\documentclass[12pt,preprint]{aastex}
\usepackage{gensymb}
\usepackage{amsmath}

\usepackage[pdftex]{hyperref}

\usepackage{mathtools}

\usepackage[skip=0pt,singlelinecheck=false,font=large]{caption}
\slugcomment{Draft of \today}

\shorttitle{The Arches Region}
\shortauthors{Hankins et al.}

\usepackage{graphicx}
\usepackage{subfigure}
\usepackage{rotating}

\usepackage[section] {placeins}

\newcommand{\angstrom}{\mbox{\normalfont\AA}}

\usepackage[labelsep=none]{caption}

\begin{document}

\title{An Infrared Study of the Dust Properties and Geometry of the Arched Filaments HII Region with SOFIA/FORCAST}

\author{M. J. Hankins$^1$, R. M. Lau$^{2}$, M. R. Morris$^3$, T. L. Herter$^1$}

\altaffiltext{1}{Astronomy Department, 202 Space Sciences Building, Cornell University, Ithaca, NY 14853-6801, USA}
\altaffiltext{2}{Jet Propulsion Laboratory, California Institute of Technology, 4800 Oak Grove Drive, Pasadena, CA, 91109-8099, USA}
\altaffiltext{3}{Department of Physics and Astronomy, University of California, Los Angeles, CA 90095-1547, USA}

\begin{abstract}

Massive stellar clusters provide radiation ($\mathrm{\sim 10^7-10^8~L_{\odot}}$) and winds ($\mathrm{\sim 1000~km/s}$) that act to heat dust and shape their surrounding environment. In this paper, the Arched Filaments in the Galactic center were studied to better understand the influence of the Arches cluster on its nearby interstellar medium (ISM). The Arched Filaments were observed with the Faint Object InfraRed CAMera for the SOFIA Telescope (FORCAST) at 19.7, 25.2, 31.5, and 37.1  $\mu$m. Color-temperature maps of the region created with the 25.2 and 37.1 $\mu$m data reveal relatively uniform dust temperatures (70-100 K) over the extent of the filaments ($\sim 25$ pc). Distances between the cluster and the filaments were calculated assuming equilibrium heating of standard size ISM dust grains ($\sim$0.1 $\mu$m). The distances inferred by this method are in conflict with the projected distance between the filaments and the cluster, although this inconsistency can be explained if the characteristic grain size in the filaments is smaller ($\sim$0.01 $\mu$m) than typical values. DustEM models of selected locations within the filaments show evidence of depleted abundances of polycyclic aromatic hydrocarbons (PAHs) by factors of $\sim$1.6-10 by mass compared to the diffuse ISM. The evidence for both PAH depletion and a smaller characteristic grain size points to processing of the ISM within the filaments. We argue that the eroding of dust grains within the filaments is not likely attributable to the radiation or winds from the Arches cluster, but may be related to the physical conditions in the Galactic center.

\end{abstract}

\keywords{Galactic Center, Arched Filaments, Arches Cluster}

\section{Introduction}

Perhaps the most striking large-scale feature near the center of the Galaxy at infrared and radio wavelengths is the collection of quasi-linear features called the Arched Filaments. The filaments are thought to be the heated and ionized edges of molecular clouds which give the appearance of large `arch-like' formations due to the viewing geometry of the system (Lang, Goss, \& Morris 2001; hereafter LGM2001). The filaments are located just $\sim$30 pc in projection from Sgr A*, assuming a distance of 8 kpc to the Galactic center (Reid 1993), and extend over a projected distance of $\sim$25 pc. There are two main sub-structures in the filaments which are referred to as the eastern and western filament (Figure 1). The Arched Filaments were first detected with radio observations near Sgr A* by Yusef-Zadeh, Morris, \& Chance (1984), who noted a large arc-like formation to the northeast of Sgr A*. Subsequent studies of this region have shown a complex and detailed structure comprised of thermal and non-thermal components (Yusef-Zadeh \& Morris 1987, Morris \& Yusef-Zadeh 1989). Studies of the velocity structure of the thermal filaments in ionized and molecular gas show that the velocities vary considerably along and across the filaments ($\sim$-20 to -50 km/s), indicating that each filament is likely composed of multiple components (Serabyn \& G\"usten 1987; hereafter SG1987; LGM2001; Lang, Goss \& Morris 2002; hereafter LGM2002).

The complex geometry presented by the filaments has been a major impediment for understanding the nature of the region. Early studies of emission lines from ionized gas in the region suggested that the radiation field throughout the filaments is very uniform, which is highly unusual over such large scales (Yusef-Zadeh 1986; Morris and Yusef-Zadeh 1989). The source of the heating and ionizing radiation for the filaments was unknown at the time, which made the interpretation more difficult. The discovery of the 2-3 Myr old, $L_{bol}\sim10^{7.8} \mathrm{L_{\odot}}$ ``Arches'' star cluster (Cotera et al. 1996; Figer et al. 1999b, Stolte et al. 2002) just to the east of the filaments ($\sim$1') provided a likely heating and ionization source. However, the uniformity of the ionization suggests the filaments are much farther away from the cluster than they appear on the sky. This idea was explored by LGM2001, who found that the observed ionization of the region can be explained by the Arches cluster even if the cluster is located a moderate distance away ($\sim 10-20$ pc). 

While the ionized and molecular gas in the filaments have been studied in detail in previous works (e.g., LGM2001, LGM2002, SG1987), the thermal dust emission in the region has not received as much attention (e.g., Davidson et al. 1994). Examining the dust emission of the filaments can provide additional diagnostics for studying the heating source and the geometry of the region. To investigate the thermal dust emission of the filaments, we have performed observations of the region at 19.7, 25.2, 31.5, and 37.1 $\mu$m using the Faint Object InfraRed CAMera for the SOFIA Telescope (FORCAST). The FORCAST maps present a significant improvement in spatial resolution (FWHM$\sim$3.2-3.8") over previous maps of the region at comparable wavelengths\footnote{e.g., the MSX 21 $\mu$m observations of the Galactic plane (Price et al. 2001) or the Spitzer/MIPS 24 $\mu$m observations (Carey et al. 2009) of the Galactic center where the Arched Filaments badly saturated (Hinz et al. 2009).}, allowing for detailed study of the warm dust morphology of the filaments. The goal of this investigation is to provide a detailed analysis of the dust emission from the Arched Filaments to better understand their geometry and thermal structure. The analysis presented in this paper also focuses on studying properties of the dust in the filaments. Since the filaments appear to be relatively close to the Arches cluster, it is reasonable to think that interactions with the cluster have significantly influenced the surrounding dust properties and morphology. However, the ISM throughout the Galactic center region is influenced by a variety of different processes such as frequent supernova shocks, powerful stellar winds, cloud collisions, and a strong tidal field (Morris \& Serabyn 1996), which makes it challenging to understand the contributions of individual processes.

The organization of this paper is presented as follows: First, details for the observations presented in this paper and archival data that has been used in the analysis are provided in \S2.1 and \S2.2. The method used to correct for interstellar extinction is described in \S3. The results of this paper are presented in \S4, including a study of the morphology (\S4.1), dust temperatures (\S4.2), infrared luminosity (\S4.3), and dust mass (\S4.4) of the Arched Filaments. In \S4.5, DustEM is used to generate models of the dust emission of selected locations within the filaments to study variations in the dust composition. This is followed by a discussion of interesting regions within the filaments (\S5.1), factors which might play a role in shaping the morphology of the filaments (\S5.2), and grain processing mechanisms which may be acting on dust in the region (\S5.3).

\section{Observations and Data Reduction}

\subsection{SOFIA/FORCAST}

Observations of the Arched Filaments were made using FORCAST on the 2.5 m telescope aboard the Stratospheric Observatory for Infrared Astronomy (SOFIA; Herter et al. 2012). FORCAST is a $256 \times 256$ pixel dual-channel, wide-field mid-infrared camera with a plate scale of $0.768''$ per pixel and field of view of $3.4'\,\times\,3.2'$. The two channels consist of a short-wavelength camera (SWC) operating at $5 - 25~\mu\mathrm{m}$ and a long-wavelength camera (LWC) operating at $28 - 40~\mu\mathrm{m}$. An internal dichroic beam-splitter enables simultaneous observation from both short-wavelength and long-wavelength cameras, while a series of bandpass filters is used to select specific wavelengths.

FORCAST observations of the Arched Filaments were made during SOFIA Cycles 3 \& 4. Details of the flights over both cycles can be found in Table 1. Since the filaments are much larger ($\sim8'\times5'$) than the FORCAST field of view, the region was mapped using six different pointings to provide a complete coverage. The filaments were observed with the 19.7, 25.2, 31.5, and 37.1 $\mu$m filters. To increase the efficiency of mapping in the region, fields were simultaneously observed in the short-wavelength and long-wavelength channels by using the dichroic beamsplitter. Additional observations using only the long-wavelength channel at 37.1 $\mu$m were made to improve the signal-to-noise ratio of these maps. Figure 2 shows the locations of the fields observed, along with the naming conventions used in this paper.

Chopping and nodding were used to remove the sky and telescope thermal backgrounds. An asymmetric chop pattern with the source on the telescope axis, which eliminated optical aberrations (coma) on the source. The off-source chop fields (regions of low mid-infrared Galactic emission) were selected from the Midcourse Space Experiment (MSX) $21~\mu$m image of the Galactic Center and are shown in Figure 2. The frequency of the chop throw was $\sim 4$ Hz. 

A five-point dither pattern was used to aid in the removal of bad pixels and to mitigate response variations.
The integration time at each dither position was $\sim 20$ sec. The pointing accuracy of the telescope is $\sim0.5"$. The quality of the images is consistent with
near-diffraction-limited imaging at $19.7 - 37.1~\mu$m; the full width at half maximum (FWHM) of the point spread function (PSF) was $3.2''$ at $19.7~\mu$m, $3.4''$ at $25.2~\mu$m, $3.6''$ at $31.5~\mu$m, and $3.8''$ at 37.1 $\mu$m.

The acquired images were reduced and combined at each wavelength according to the pipeline steps described
in Herter et al. (2013). Calibration was determined from the average response taken over the OC1-B Flight Series, adjusted to those of a flat spectrum ($\nu F_\nu = \rm{constant}$) source. The $3\sigma$ uncertainty in the calibration is estimated to be $\pm20\%$. The RMS noise per pixel was $8.3$ mJy at $19.7~\mu$m, $14.9$ mJy at $25.2~\mu$m, $34.2$ mJy at $31.5~\mu$m, and $37.5$ mJy at 37.1 $\mu$m.

The FORCAST images of the different fields were combined into a single mosaic of the region. The relative positioning for each field was determined by using cross-correlation analysis of the fields with the HST/Paschen-$\mathrm{\alpha}$ (Wang et al. 2010) and Spitzer/IRAC 8 $\mu$m (Churchwell et al. 2009) maps, which provided a rough location for each field. The locations of fields containing point sources were further refined using the positions of the sources compared with the 8 $\mu$m map. Finally, the combined maps were deconvolved using the Richardson-Lucy algorithm (Richardson 1972, Lucy 1974) with Gaussian kernel used to simulate the PSF. The final maps of the region were then convolved back to a uniform beam size of $3.4''$.

\subsection{Data from Other Sources}

Archival data from a variety of sources were incorporated into this analysis of the Arched Filaments. The HST Paschen-$\mathrm{\alpha}$ Survey of the Galactic center (Wang et al. 2010) was used to examine the spatial distribution of ionized gas in the region, and GLIMPSE Spitzer/IRAC survey (Churchwell et al. 2009) observations were incorporated into the analysis of the filaments to expand the wavelength coverage for spectral energy distributions (SEDs) of the region. The mid-infrared emission sampled by the IRAC bands is important for characterizing the relative abundance of PAHs in the region. Additionally, the 8 $\mu$m IRAC maps were used to study the spatial distribution of PAHs and very small grains (VSG) in the region. To aid in characterizing the far-infrared emission of the filaments, we referenced maps of the region taken at 70 and 160 $\mu$m with Herschel/PACS that were observed as a part of the Herschel Hi-GAL survey (Molinari et al. 2010). 

In addition to the available maps of the region, there are several locations within the filaments where infrared spectra (R $\sim$ 600) were taken with Spitzer/IRS (Program 3295). These spectra were previously presented in Simpson et al. (2007). Calibrated spectra were obtained from the Spitzer Heritage Archive. The short-wavelength (SW) and long-wavelength (LW) spectra were combined into a single spectrum after correcting for differences in the aperture sizes. As a check, the corrected, processed spectra were compared with the continuum values quoted in Simpson et al. (2007) and were found to be within the quoted error. Finally, the SW spectrum was divided by an additional factor of 1.3 to improve the agreement in continuum levels between the LW and SW data. To mitigate any issues that may be caused by scaling the SW data, these data were not used to fit any parameters unless the added spectral coverage between the IRAC bands and FORCAST bands ($\sim10-20~\mu$m) was needed. The LW IRS spectra were used to provide an independent check of the FORCAST flux calibration.

\section{Interstellar Extinction}

Objects near the Galactic center are known to suffer significant extinction ($A_V\sim30$) due to large column densities of gas and dust along the line of sight (e.g., Cardelli et al. 1989). In this work, the extinction law of Fritz et al. (2011) was adopted to deredden the observed fluxes of the Arched Filaments. The Fritz law was derived from hydrogen recombination line observations of Sgr A West at 1-19 $\mu$m with the Short Wave Spectrometer (SWS) instrument on the Infrared Space Observatory (ISO) and the Spectrograph for INtegral Field Observations in the Near-Infrared (SINFONI) on the Very Large Telescope (VLT).

In correcting the maps of the filaments, extinction values consistent with extinction toward the Galactic center were adopted ($\mathrm{A_{k_s}}=2.42$; Fritz et al. 2011). Since the line of sight extinction correction may vary as a function of Galactic latitude, changes to the scaling of the extinction curve were considered in the analysis. The available Spitzer/IRS spectra at different locations in the filaments may provide some estimate of variations in the extinction with Galactic latitude from differences in the depth of the 9.7 $\mu$m silicate absorption feature. Spatial variations in the depth of the silicate feature could indicate variable line-of-sight extinction but could also be caused by spatially variable silicate emission along the line-of-sight, which makes this assessment less certain. Nevertheless, the relative changes to the silicate feature depth were used to estimate variations in the line-of-sight extinction.

The depth of the silicate feature in the spectra shows only small variations ($\sim$10-15\%) for most of the locations in the filaments observed by Simpson et al. (2007), indicating little change in extinction with Galactic latitude for the region. This supports our use of a uniform extinction correction over the entire region. The only location showing a substantial decrease ($\sim50\%$) in the silicate feature depth was at the highest galactic latitude in the western filament. For this location, the extinction correction was allowed to vary in the DustEM models of the region to better understand how well constrained the various model parameters are. These results can be found in \S 4.5.

\section{Results and Analysis}

\subsection{Morphology of the Arched Filaments}

Observations with FORCAST reveal the warm dust ($\sim 100$ K) of the Arched Filaments with better spatial resolution ($\sim3.4"$) than previous observations at comparable wavelengths. To study the spatial distribution of dust emission in the region, contours from the 37.1 $\mu$m maps have been overlaid on maps of the region at 1.87 $\mu$m (Paschen-$\alpha$), 8 $\mu$m, 70 $\mu$m, and 160 $\mu$m (Figure 3). Mid-infrared continuum emission often originates from dust inside HII regions as well as dust in photon dominated regions (PDRs). Because of this, it can be useful to plot the 19.7 and 37.1 $\mu$m contours separately since they could be dominated by different temperature dust components (e.g., Salgado et al. 2012); however, the spatial distributions of the 19.7 and 37.1 $\mu$m emission show relatively little difference so only the 37.1 $\mu$m contours are plotted in Figure 3 for clarity.

A comparison of the 37.1 $\mu$m emission with the Paschen-$\alpha$ emission shows good spatial correlation. Interestingly, the 37.1 $\mu$m emission does not correlate spatially with the 8 $\mu$m emission in parts of the western filament. There appears to be additional 8 $\mu$m emission to the west of the observed extent of this filament at 37 $\mu$m. The 8 $\mu$m map traces emission from PAHs and very small grains (VSGs) which typically originates relatively deep within a cloud but can also trace regular-sized grains if they are sufficiently hot ($\sim$250 K). If the PAH emission and warm dust emission traced by 37 $\mu$m emission originate at the same location in the cloud, we would expect these two components to follow one another. The lack of 37 $\mu$m emission to the west of the western filament cannot be accounted for by the sensitivity threshold of the 37 $\mu$m data, although the the lack of extension of the filament to the north at 37 $\mu$m is likely attributable to this. Overall, the 70 $\mu$m emission in this region, which can originate from colder dust than the 37 $\mu$m emission, traces the 8 $\mu$m emission, indicating that the PAH emission may originate from a location similar to that of the colder 70 micron $\mu$m component and not the warmer 37 $\mu$m component.

To compare the FORCAST maps with the cold dust components from the region, Figure 3 also shows the 37.1 $\mu$m contours overlaid on the Herschel/PACS 70 and 160 $\mu$m maps. In the filaments, the 37.1 $\mu$m emission generally traces the 70 $\mu$m emission; however, there is an interesting change in the morphology of the region at 160 $\mu$m. The filaments are no longer as prominent at this wavelength and are difficult to differentiate from other emission in the region. There are a few strong peaks of dust emission at 160 $\mu$m just to the east of the eastern filament. A few of these peaks appear to be coincident with the E1 filament; however, the brightest emission peak at 160 $\mu$m does not appear to correspond to any features from the filaments seen at 37.1, 70 $\mu$m or in the Paschen-$\alpha$ maps. It is somewhat curious that the filaments themselves seem to lack significant 160 $\mu$m emission. The 160 $\mu$m emission could originate from relatively deep, cold parts of the filaments clouds that are displaced in projection or the emission could originate from unrelated clouds along the line of sight.

Line cuts through the filaments were used to look for spatial variations in the dust emission at 8, 25.2, 37.1, and 70 $\mu$m. These are able to show variations on scales ($\sim 8''$) that can be more difficult to distinguish in the contour plots (Figure 4). The different wavelengths that are plotted trace warm dust ($\sim$100 K at 25.2 and 37.1 $\mu$m), PAHs and VSGs (at 8 $\mu$m), and cooler dust ($\sim$40 K at 70 $\mu$m). In a few of the lines (1 \& 5), the peaks of all four wavelengths are coincident but the majority of the line cuts show more complex profiles. For example, line cuts 2 and 8 show the peak of 8 $\mu$m emission significantly offset from the peak emission at 25 and 37 $\mu$m. Line 4 shows an interesting double-peaked structure in the 8 $\mu$m emission, with one peak coincident with the 25 and 37 $\mu$m peak and the other offset by $\sim$20". The spatial variations of the emission in some of these regions are quite interesting and can be explained by either optical depth variations or differences in distance between the clouds and their heating source. Warm dust, which can be traced by all four components, must be relatively close to a heating source in order to maintain its temperature ($\sim$100 K). Whereas cooler dust ($\sim$40 K), traced primarily by the 70 $\mu$m emission, must be sufficiently far from a heating source or is shielded from more intense radiation by intervening dust extinction. In the following sections, we examine many of these regions in further detail.

\subsection{Color-Temperature Map and Heating Analysis}

The dust temperatures in the region were estimated by solving for color-temperatures using the emission from the 25.2 and 37.1 $\mu$m maps. This was done by expressing the emission from the dust as $F_{\nu}\propto \nu^{\beta} B_{\nu}(T_D)$, which assumes the emission is optically thin with an emissivity of the form $\nu^{\beta}$. In this expression we define $T_D$ as the color temperature of the dust, and we adopt an index of $\beta=2$, which is consistent with standard interstellar grains (Draine 2003). The color-temperature map of the dust in the region is shown in Figure 5. The filaments show a remarkable temperature uniformity ($\sim$ 70-100 K) over fairly large spatial scales ($\sim$25 pc in projection). The uniform temperature observed over the eastern filament is particularly interesting since locations along this filament have projected distances which are only a few parsecs from the Arches cluster at closest and $\sim$10 pc at furthest. This strongly suggests that the true distance between the cluster and the closest points of the filament must be further away than they appear.

The true distance between the filaments and the Arches cluster can be investigated using calculations for the equilibrium heating of dust grains. For a dust grain in thermal equilibrium with an incident radiation field, the temperature is determined by the optical properties of the grain and the local radiation field. If the dust is heated by a source of known luminosity, knowledge of dust temperature can be used to estimate the distance, $d$, of the dust from its heating source:

\begin{equation}
d = \Big( \frac{L}{16 \pi \sigma T_{d}^{4+\beta} } \frac{Q_{uv}}{Q_{d}} \Big)^{1/2}
\end{equation}

\noindent Where $L$ is the luminosity of the heating source, $T_d$ is the observed dust temperature, $Q_{UV}$ is the grain absorption cross section averaged over the spectrum of the heating source, $Q_d$ is the frequency-averaged dust emission efficiency given by $1.3 \times 10^{-6} (a/0.1)~\mathrm{K}^{-\beta}$ (Draine 2011). The parameter $a$ is the grain size in $\mu$m and $\beta$ represents the power-law dependence of $Q_d$ on temperature in units of K. As stated previously, we have taken $\beta=2$ (Draine 2003).

The left panel of Figure 6 shows the calculated equilibrium heating distances between the filaments and the Arches cluster, assuming a total luminosity of $L_{cluster}=10^{7.8} L_{\odot}$ (Figer et al. 1999b) and 0.1 $\mu$m silicate dust grains with $Q_{UV}=1.04$. The figure also shows the location of the Arches cluster along with circles indicating projected distances from the cluster for reference. From the equilibrium heating analysis, it is clear that the obtained distances are inconsistent with the assumption that the Arches cluster is the heating source, since they would be shorter than their projection into the sky plane. For example, boxes 1 and 2 have average projected distances of $\sim$7.5 and $\sim$15 pc from the Arches cluster, while the distances implied by the heating of 0.1 $\mu$m dust dust grains in these regions are only $3.4^{+2.3}_{-1.1}$ and $5.9^{+1.9}_{-2.1}$ pc, respectively. In this analysis, we have factored in uncertainties in the color-temperatures ($\sim$15\%) attributable to uncertainties in flux calibration ($\sim$20\%). The inconsistencies in the projected distances and the distances determined assuming equilibrium heating call into question the assumption that the Arches cluster is completely responsible for heating the region. While there may be additional uncertainties in the cluster luminosity ($\sim$60\%; Figer et al. 1999b, Figer et al. 2005), they alone are not sufficient to affect this result as matching the projected distances would require the cluster be at least $\sim$4 times more luminous than we have assumed.

To explain the heating of the filaments, either our initial assumptions about the heating source(s) or the grain properties must be incorrect. The Arches cluster will dominate the heating of the grains unless there are heretofore unknown local sources. If the heating were internal it would at least require O/B stars having total luminosities within $\sim10^{3.7-4.5} L_{\odot}$ per 3.4'' beam embedded within the filament to produce relative uniformity of the observed temperatures. This is inconsistent with near infrared observations which do not show a sufficient number of candidate sources coincident with the filaments (Dong et al. 2012). Furthermore, it is already well established that the Arches cluster is the primary source of ionizing radiation for the filaments (LGM2001) which supports the idea that the cluster is also likely the dominant source of dust heating. Since local heating sources are an unlikely solution, we next consider alternative dust properties. Using 0.01 $\mu$m silicate grains rather than 0.1 $\mu$m grains results in distances inferred by the equilibrium heating calculations that agree well with the projected distances from the Arches cluster (Figure 6, right panel). Additionally, the distances implied by 0.01 $\mu$m grains are mostly consistent with the distances to parts of the filaments derived by studying the ionization of the filaments (LGM2001). The small characteristic size of dust in the filaments implied by this analysis ($\sim$0.01 $\mu$m) might be indicative of significant processing of the ISM in the region. We will return to this point in more detail in \S 5.2. 

\subsection{Infrared Luminosity and Dust-Covering Fraction}

To better understand the properties of the filaments, the infrared luminosity of the filaments was calculated. This was done by fitting a simple dust emission model to each pixel in the maps of the filaments. The wavelengths used to construct the fitted SEDs consisted of the four FORCAST bands at 19.7, 25.2, 31.5, and 37.1 $\mu$m and two Herschel/PACS bands at 70 and 160 $\mu$m. The dust emission models used in the fitting were created by the radiative transfer code DUSTY (Nenkova et al. 2000) using a simple plane-parallel geometry provided with the code. A set of DUSTY models was created by changing the inner boundary temperature of the plane from 50 to 120 K in increments of 5 K. The luminosity of each pixel was derived by integrating the flux from the dust model that best corresponds to its 25/37 color-temperature. To ensure the goodness of the model fits, maps of the emission based on the models were generated to compare with the observed maps of the region. Relative to the models, the observed data revealed a far-infrared excess at a few locations in the region.

The need for a secondary cold dust component to match the continuum emission from the filaments was already noted by Kaneda et al. (2012), who used AKARI observations of the Arched Filaments and the Sickle HII region to study the far-infrared fine-structure line emission of these regions. For our models of the filaments, an additional 20 K DUSTY model was used to fit the observed far-IR fluxes that exceeded the prediction of DUSTY models based on the 25/37 color-temperatures. Figure 7 shows the spatial distribution of the infrared luminosity for the warm dust component (left) and cold dust component (right). In total, the infrared luminosity of the warm component is $L_{IR}=8.2 \pm 2.5 \times 10^6 L_\odot$. In comparison, the total luminosity of the cold component is only $\sim10\%$ of that of the warm component.

Given the infrared luminosity of the filaments, the dust-covering fraction can be calculated. The dust-covering fraction is defined as the fraction of the total solid angle around the Arches Cluster that is covered by dust, which is equivalent to the fraction of the luminosity from the cluster that is absorbed and re-radiated in the infrared (i.e., $L_{IR}/L_{cluster}$, which assumes that the dust clouds are optically thick and therefore absorb all of the Arches cluster starlight that hits them). Comparing the total infrared luminosity from the warm dust in the filaments to the total luminosity of the Arches cluster yields a value for the dust-covering fraction of $0.13 \pm 0.04$. The luminosity of the cold dust has been omitted because it is only small fraction of the total infrared luminosity and not all of the cold dust emission may be associated with the filaments or with the Arches cluster. In estimating the uncertainties for the dust-covering fraction, only the uncertainties in the infrared luminosity were considered. The Additional uncertainty in the luminosity of the Arches cluster was not included.

Utilizing the luminosity-derived dust-covering fraction, we can make a rough estimate of the surface area subtended by the filaments from the perspective of the Arches cluster to compare with their projected surface areas. If the infrared emission from each filament is confined to a plane on the surface of the molecular cloud(s) which is nominally optically thick to the photons heating the dust (so that all the radiation in the direction of the filaments is absorbed), then the illuminated surface area of the filaments is related to the luminosity-derived covering fraction via:
 
\begin{equation}
\frac{L_{IR}}{L_{cluster}}=x\frac{l~w}{4 \pi d^2}
\end{equation}
 
\noindent Where $l$ and $w$ are the projected length (larger dimension, $\sim$16 pc) and width (shorter dimension, $\sim$3.2 pc) of each of the filaments, $d$ is the distance between the filaments and the cluster, and $x$ is a projection factor to account for differences in the apparent surface area of the filaments (based on their 2D dimensions on the sky) and the surface area presented by the filaments from the perspective of the Arches cluster. We adopt average distances between the filaments and the cluster consistent with our equilibrium heating calculation for 0.01 $\mu$m grains ($d_{East}=8.6^{+2.9}_{-2.0}$ pc and $d_{West}=12.1^{+2.8}_{-2.0}$ pc). Solving for x, we find the projected surface areas of the eastern and western filaments are factors of $\sim$2.2, and $\sim$2.1 smaller than the surface area of the filaments from the perspective of the Arches cluster. If the filaments are inclined in such a fashion that their length and width from the perspective of the Arches cluster are both larger by a factor of $\sqrt2$ (i.e., a 45$\degree$ angle), this would roughly account for the additional surface area. However, there are degeneracies in the inclination angles of the length and width of the filaments that could also satisfy the constraints for the dust-covering fraction of the filaments.

\subsection{Optical Depth Map \& Observed Dust Mass}

For optically thin dust emission, the optical depth can be expressed as:

\begin{equation}
\tau_{\lambda}=\frac{F_{\lambda}}{\Omega_p B_{\nu}(T_d)}
\end{equation}

\noindent Where $F_{\lambda}$ is the flux per pixel, $\Omega_p$ is the solid angle subtended by each pixel, and  $B_{\nu}(T_d)$ is the Planck function at the dust temperature, $T_d$. A map of the optical depth of the filaments at 37 $\mu$m was created using the 25/37 color-temperature map and the dust emission intensity from the 37 $\mu$m map (Figure 8). The opacity map of the region shows optical depths ranging from $\tau_{37}\sim10^{-3}$ to $10^{-2}$, with the larger values tracing the bright emitting regions at 37 $\mu$m. The largest opacities observed in the map occur near the base of the eastern filament, corresponding to the region near G0.10+0.02.

The optical depth map was used to obtain a measurement of the observed mass in the region. The optical depth, $\tau_{\nu}$, at a frequency, $\nu$, can be converted to dust mass per pixel, $M_d$, via:

\begin{equation}
M_d=\frac{4/3~\Omega_p~d^2~ a~\rho_b}{Q_{abs}(a,\nu)} ~\tau_{\nu}
\end{equation}

Where $\Omega_p$ is the solid angle subtended by a pixel, $d$ is the distance to the Galactic center, $a$ is the grain size (assumed to be 0.01 $\mu$m), $\rho_b$ is the bulk density of grains (3.5 g/cm$^{-3}$; Draine 2011), and $Q_{abs}$ is the grain absorption coefficient. The total observed dust mass of the filaments is $6.8 \pm 2.4~\mathrm{M{_{\odot}}}$. Any change to the grain size in this calculation results in a negligible change in dust mass since $Q_{abs}\propto a$ for $0.0005~\mu$m $\lesssim a \lesssim$ $1.0~\mu$m.

To compare the observed dust mass from the FORCAST observations with other measures of the mass in the filaments, it is useful to convert the dust mass to gas mass. Adopting a gas-to-dust ratio of $\sim$100 (Draine et al. 2007), this results in a total gas mass of $6.8 \pm 2.4 \times10^2~\mathrm{M_{\odot}}$. The inferred gas mass agrees well with the observed ionized gas mass associated with the Arched Filaments HII region, $M_{HII}\sim6.7 \times 10^2~\mathrm{M_{\odot}}$ (LGM2001), and is much smaller than the molecular gas mass of the filaments, $M=6 \pm 3 \times 10^5 ~\mathrm{M_{\odot}}$ (SG1987). The small observed total mass demonstrates that the FORCAST observations sample the material at the ionized edges of the clouds and do not probe more deeply into the molecular region which is accessible at longer wavelengths. If the observed dust mass in this work is representative of the total mass of dust in the HII region, it would imply a gas-to-dust ratio of $98^{+53}_{-26}$ which is consistent with the value of 100 that was initially assumed. One might wonder whether the canonical gas-to-dust ratio of 100 is appropriate for a region in which the dust grains are an order of magnitude smaller in radius than typical interstellar grains. However, some of the dust emission may also be coming from the PDR as well as the HII region, which would result in the gas-to-dust ratio potentially being much larger than our estimate.

\subsection{DustEM Models}

\subsubsection{Model Parameters}

DustEM (Compi{\`e}gne et al. 2011) was used to model the dust emission from several separate locations within the filaments to determine whether any differences in the radiation field or grain abundances exist between the regions. DustEM incorporates transient heating of grains and is thus able to model the emission from stochastically heated very small grains, including PAHs, which is not possible with DUSTY. The selected regions, shown in Figure 9, consist of locations that were observed with Spitzer/IRS by Simpson et al. (2007), denoted in blue, along with other locations spread through the eastern and western filament, denoted in red. 

SEDs for the regions were constructed using data from the four Spitzer/IRAC bands at 3.6, 4.5, 5.8, and 8.0 $\mu$m, the four SOFIA/FORCAST bands at 19.7, 25.2, 31.5, and 37.1 $\mu$m, and the Herschel/PACS 70 $\mu$m data.\footnote{Since the 160$\mu$m data appears to be dominated by the possibly unrelated far-infrared excess, it is not used in the fitting process.} The fluxes for the SEDs were taken from an $18"\times18"$ extraction region at each position. Before extracting the fluxes, the FORCAST and PACS data were brought to the same spatial resolution by deconvolving the maps and convolving them back to a uniform spatial resolution ($\sim$8"). However, the IRAC data were left at their native resolution because of the contamination by many point sources in the region. The extraction region for the IRAC data was sub-sampled in several smaller regions to provide a better estimate of the diffuse dust emission by avoiding point sources in each region.

For the DustEM models, we adopt dust properties and abundances for the diffuse ISM from Compi{\`e}gne et al. (2011). This model consists of four grain species with the following abundances by mass (in terms of $M/M_H$): PAHs ($Y_{PAH}=7.8\times10^{-4}$), small amorphous carbon (SamC; $Y_{SamC}=1.65\times10^{-4}$), large amorphous carbon (LamC; $Y_{LamC}=1.45\times10^{-4}$), and astronomical silicates (aSil; $Y_{aSil}=7.8\times10^{-3}$). The abundances of PAH and SamC species were allowed to vary from these initial values to achieve better fits to the SEDs. The relative abundances of the aSil and LamC components were left fixed in the models since the emission of these components is similar and it is difficult to distinguish their individual contributions.

The input radiation field for the DustEM models was a combination of radiation from the Arches cluster and the interstellar radiation field (Mathis et al. 1983). Radiation from the Arches cluster was simulated using the stellar population synthesis code Starburst99 (Leitherer et al. 1999). The parameters adopted for the Arches cluster are listed in Table 2. Based on the Starburst99 model, the overall luminosity of the simulated cluster was $L_{model}=10^{7.5} L_{\odot}$. The Starbust99 model was then scaled to match the reported luminosity of the Arches cluster ($L_{model}=10^{7.8} L_{\odot}$; Figer et al. 1999b). To emulate the radiation field experienced by the dust in the region, the radiation field was scaled with a geometric dilution factor to account for the distance between the cluster and the dust. Modeling the radiation field in this manner makes it possible to estimate the distance between the dust in the filaments and their heating source, similar to what was done in \S4.2.

For each model, four main parameters were fit: the distance of the dust from the cluster, the factors by which the abundances of SamC dust grains and PAHs differ from the adopted ISM values (denoted as $X_{PAH}$ and $X_{SamC}$), and the overall scaling of the model to the observed infrared emission (which is related to the dust column density and infrared luminosity of the model). In addition to these parameters, we also consider the effect of using different size silicate grains, which can change the fitted distance between the filaments and their heating source. Two different silicate dust grain size distributions were adopted in the models: one with a grain size distribution characterized by a power law with $dn/da \propto a^{\alpha}$ and $\alpha=-3.4, a_{min}=4\times10^{-3} ~\mu$m, and $a_{max}= 2~\mu$m, and the other composed of uniform $a=0.01~\mu$m grains, consistent with the characteristic grain size of the filaments determined in \S 4.2. The best-fit model parameters were found by constructing a grid of models using the various parameter values and calculating the reduced $\chi^2$ for each model. 

\subsubsection{Model Results}

Following the procedure outlined above, the best-fit models for each region were found. The best-fit parameters and uncertainties for both grain size distributions are listed in Table 3, and the best-fit models for the uniform grain size distribution are shown in Figures 9 and 10. As expected, the distances between the filaments and the Arches cluster fitted using $a=0.01~\mu$m grains agree well with the results from the equilibrium heating calculations. The distances inferred from the power-law models are too small compared to the projected distances for the majority of the locations. In Table 3, the best-fit values for both types of models are presented to show other differences in the models, such as the relative changes in the PAH and SamC abundances. However, we favor the models using a uniform silicate grain size of $a=0.01 \mu$m. The silicate grains in the filaments are actually likely to have a size distribution which lacks larger dust grains and has a characteristic size of $\sim0.01~\mu$m. We have adopted the uniform size distribution for simplicity. For these models, the best-fit parameters show evidence for variations in the PAH abundances in different regions of the filaments. For example, region 34 appears the most depleted in PAHs, with a relative abundance by mass $\sim10$ times smaller than in the diffuse ISM, while region W2 has the highest PAH abundance which is only $\sim$1.6 times smaller than the diffuse ISM abundance.

The models also indicate variations in the abundance of the SamC component between some of the regions. Tracking such variations for all the regions is more difficult because the continuum emission from the SamC component consistently peaks between 10-20 $\mu$m. This region of the spectrum is not covered by any of the broadband filters used in our analysis but is sampled with Spitzer/IRS SW spectra which is only available for five of the locations. Therefore, it is only possible to constrain the abundances of the SamC component in these five regions. For all of these locations, it is clear that the abundance of the SamC component needs to be increased over the standard abundance to better match the continuum. To demonstrate this, Figure 9 shows two fits to the SED of location 30: one with an increased abundance of SamC dust to show that it is a good fit to the IRS spectrum (model 30) and a second model which has the standard abundance of SamC grains which fits very poorly to the IRS spectrum (model 30*). The enhancements of SamC abundance listed in Table 3 indicate the minimum increase in abundance needed to match the Spitzer/IRS spectrum. At longer wavelengths ($>20\mu$m), the emission from the SamC component behaves similarly to the emission from the silicate grains, making it difficult to distinguish these dust components. For this reason, we adopt the lower limit to the increase in the SamC abundance for the models.

To further test the robustness of the model parameters, the extinction correction was allowed to vary for regions which might suffer from less line-of-sight extinction. Of the different locations with Spitzer/IRS spectra, region 35 shows a much shallower silicate absorption feature, which could indicate less extinction toward this region. A second DustEM model of this location was created using the Fritz extinction law scaled proportionally to the observed change in the depth of the silicate feature in this region (corresponding to $A_{K_s}=1.25$). The best-fit model parameters for this region, denoted by 35$^{\dagger}$, were found using the same methods as for the other regions. The main changes in the best-fit parameters in this region are an increased distance from the Arches cluster and a slight decrease in the PAH abundance, which is not statistically significant. Other than region 35, none of the other locations with IRS spectra show any indication of decreased line-of-sight extinction.

\section{Discussion}

\subsection{Notable Regions Within the Filaments}

Having already addressed some of the large-scale morphological results, we begin here by comparing some of the interesting locations within the filaments. The region near G0.10+0.02 (see Figure 1) exhibits some of the highest densities and lowest temperatures in the filaments. Line cut 2, which intersects this region (Figure 4), shows a displacement between the peak of emission from warm dust at 25 and 37 $\mu$m and the 8 $\mu$m emission, with the 8 $\mu$m peak occurring near the regions of highest density that are further away from the Arches cluster. The DustEM models of this region show that the locations near the peak of the 8 $\mu$m emission (SE3) may be enriched in PAHs relative to the location of the warm dust emission peak (SE2). These properties indicate that the high column density regions exhibiting emission from cooler dust are located deep within the cloud where the PAHs have a relatively higher abundance.

An examination of the region near G0.07+0.04 (line cut 3 in Figure 4) reveals some characteristics similar to those of G0.10+0.02. The column densities and temperatures near G0.07+0.04 mimic what is observed in G0.10+0.02; however, the peaks of high column density and low temperature in this region occur close to the signal-to-noise cutoffs in the mid-infrared maps, which makes the data somewhat suspect. However, the line cut through G0.10+0.02 and the contour maps show an offset between the location of the emission peak of the warm dust (traced by the 37 $\mu$m emission) and the cold dust (traced by the 70 and 160 $\mu$m emission), which supports the idea that there is a temperature gradient in this region that is perpendicular to the ridge of warm dust. While G0.10+0.02 and G0.07+0.04 appear to have somewhat similar properties, G0.07+0.04 is one of the few locations showing substantial emission at 160 $\mu$m.

In addition to these major regions within the filaments, there are several interesting, small-scale features in the filaments. Overall, the eastern and western filaments run north-south with some degree of curvature. However, there are a few wispy features observed in the FORCAST maps which appear to run perpendicular to the filaments (we have marked two of the stronger features with arrows in Figure 1). The Paschen-$\alpha$ maps of the region, which have much higher spatial resolution ($\sim$0.25") than the FORCAST maps, easily detect these features as well. Examining individual small-scale features of the ionized gas in the filaments is quite interesting because there is a considerable amount of substructure within the filamentary HII regions. The individual, slender, threadlike features which make up the clouds may point to strong magnetic fields in the region (Morris et al. 1992). 

\subsection{Investigating the Morphology of the Filaments}

In addition to studying individual features in the filaments, it is also important to understand what mechanisms may be driving the overall morphology of the region. Given the close proximity of the filaments and the Arches cluster, one might expect that the cluster is significantly influencing the clouds in the filaments. In previous sections, it has been shown that the radiation field of the Arches cluster is likely responsible for the observed heating of the filaments, and other works have shown that the cluster is also responsible for their ionization (LGM2001). To determine whether the winds from the cluster might also have an observable effect on the filaments, we consider the ram pressure of the winds impacting the cloud compared to the turbulent pressure of the gas in the cloud. The radius of influence, $r$, of the cluster can be calculated by equating these two pressures. This gives:

\begin{equation}
r=\sqrt{\frac{\dot{M}_{cluster}~v_{cluster~wind}}{4 \pi \rho_{arches}~\Delta v_{arches}^2}}
\end{equation}

\noindent Where $\dot{M}_{cluster}$ is the mass loss rate from the cluster ($4\times10^{-4}~ \mathrm{M_{\odot}/yr}$; Yusef-Zadeh et al. 2002), $v_{cluster~wind}$ is the wind velocity of the cluster from the perspective of the cloud ($\sim$1200 km/s; Yusef-Zadeh et al. 2002), $\rho_{arches}$ is the density of the filamentary arches cloud ($\sim1.4\times10^{-20}\mathrm{~g/cm^{3}}$; SG1987),  and $\Delta v_{arches}$ is the internal velocity dispersion of the filament clouds ($\sim$30 km/s; LGM2001). These values result in a radius of influence of only 0.45 pc. Thus, the momentum of the cluster wind is too small to alter the cloud momentum in any significant way beyond a distance of $\sim$0.5 pc. This is much smaller than the projected distance between the filaments and the cluster, implying that the winds from the cluster are not responsible for shaping the filaments.

Alternatively, it has been suggested that the unusual morphology of the filaments may be driven by tidal shearing of the filaments' clouds in the gravitational potential of the Galactic center (SG1987, LGM2001). The densities derived for two bright CS J=2-1 molecular line emission peaks in the filaments ($n_{peak1}\sim5\pm3\times 10^4~\mathrm{cm^{-3}}$ and $n_{peak2}\sim2\pm1\times 10^4~\mathrm{cm^{-3}}$; SG1987) are close to the density for a cloud to be tidally disrupted at their approximate distance from Sgr A* ($n_{disrupt}\sim5\times 10^4~\mathrm{cm^{-3}}$; SG1987).\footnote{The approximate radius at which a cloud will be tidally disrupted by the gravitational potential in the Galactic center can be approximated as $R\sim75~\langle n_4 \rangle^{-0.5}$ (SG1987). Where $R$ is the radius in pc and $n_4$ is the gas density in units of $10^4~\mathrm{cm}^{-3}$.}

The observations and models presented in this paper provide additional density estimates for regions of the filaments clouds which were not observed in SG1987. For example, the region G0.10+0.02 has a dust column density of $\sim 4\times10^{-5}$ g/cm$^2$. If we assume that the depth of the cloud is approximately the same as the width of the bright emitting region of the cloud ($\sim$0.5 pc) and adopt a gas-to-dust ratio of 100 (Draine 2007), the density of the gas can be estimated $n\sim1.4\times10^3$ cm$^{-3}$. This value is close to the derived densities for other parts of the filaments from SG1987, and this density is also below the density threshold for tidal disruption at the location of the filaments. All of the locations that were modeled with DustEM have similar column density to G0.10+0.02 within a factor of $\sim$2, which would suggest that these regions have fairly comparable number densities. This result favors the scenario of the filaments' clouds being tidally sheared as they orbit around the Galactic center. However, it should be noted that our number densities could be underestimated if the depth of the clouds is smaller than we have estimated and/or the gas-to-dust ratio is larger than our adopted value.

The timescale for the filaments' clouds to be tidally sheared in the gravitational potential of the Galactic center can also be estimated. The gravitational potential of the Galactic center at the location of the Arched Filaments is not well known, although we can make a reasonable approximation assuming that the enclosed mass in the Galactic center increases linearly with distance, $M\sim R$, where $R$ is the galactocentric radius.  Under this assumption, the orbital velocity of different parts of a cloud, $v$, is constant; however, the angular velocity, $\omega$, goes as $R^{-1}$, with $R$ representing the galactocentic radius of the cloud. The tidal shearing  timescale can be estimated by the time it takes for opposite sides of the cloud to be displaced azimuthally by the cloud diameter, $2r$. This will occur at a time, $t_{shear}$, such that $R \Delta \omega t_{shear} = 2 r$ where $\Delta \omega= v(1/(R-r)-1/(R+r))$. Solving for $t_{shear}$ yields:

\begin{equation}
t_{shear}= \frac{2r} {R v ((R-r)^{-1}-(R+r)^{-1})}
\end{equation}

For the size and the distance of the filaments' clouds, $r\ll R$, which allows the expression to be reduced to $t_{shear}=R/v$. If we adopt a cloud velocity of 200 km/s and distance of 25 pc, this implies a shearing timescale of $\sim10^5$ yr. In the next section, we will show that the tidal shearing timescale is comparable to the interaction timescale of the Arches cluster and the filaments.

\subsection{Grain Processing Mechanisms in the Filaments}

In this section, we review processing mechanisms that might be able to explain the smaller characteristic grain size of the filaments.  Since the Arches cluster is a relatively massive, luminous cluster, it is reasonable to expect that it has injected significant amounts of energy into the surrounding medium over its lifetime (2-3 Myrs; Figer et al. 1999b, Figer et al. 2002). However, the filaments and the Arches cluster have not been interacting over the entire lifetime of the cluster. The relative heliocentric velocity of the stars in the cluster ($+95\pm8$ km/s; Figer et al. 2002) and the ionized gas in the filaments ($\sim$-20 to -50 km/s; Lang et al. 2002) indicate that these components are not co-moving. The observed ionized gas on the front surfaces of the clouds in the filaments indicates that the Arches cluster is on the near side of the filaments (Cotera et al. 2000). Coupled with the velocity information, this indicates that the cluster is currently moving toward the filaments. We can roughly estimate the timescale over which the cluster is able to strongly interact with the filaments. If the distance over which the interaction between the cluster and the cloud is on the order of $\sim$10 pc, which is roughly the current distance between the filaments and the cluster, then the observed radial velocity difference implies that the timescale for the interaction is only $\sim10^5$ years.

As the cluster approaches the filaments, the winds from the cluster can affect dust grains at the interfaces of clouds by sputtering material off of the grain surfaces. We can calculate the non-thermal sputtering rate, $da/dt$, due to interactions between the cluster winds and grains in the filaments:

\begin{equation}
\frac{da}{dt}=\frac{m_g}{2 \rho_b} v_w n_H A_i Y_i(E)
\end{equation}

\noindent Where $m_g$ is the mass of the grain, $\rho_b$ is the bulk density of the grain, $v_w$ is the wind velocity of the cluster (1200 km/s; Yusef-Zadeh et al. 2002), $n_H$ is the number density of hydrogen in the wind, $A_i$ is the abundance of the impinging ion, and $Y_i(E)$ is the sputtering efficiency of the ion for a given kinetic energy $E$ (Tielens et al. 1994). The relevant number densities were derived from the mass-loss rate of the Arches cluster ($4\times10^{-4}~ \mathrm{M_{\odot}/yr}$; Yusef-Zadeh et al. 2002) as $n_H=\dot{M}/(4 \pi d^2 v_w m_H)$. Where $d$ is the distance between the filaments and the cluster ($\sim$10 pc) and $m_H$ is the mass of hydrogen. These values provide a non-thermal sputtering rate of $1\times10^{-4}~\mathrm{\angstrom/yr}$, which is far too small to allow for processing of large dust grains (0.1 $\mu$m) into small grains (0.01 $\mu$m) in the available time scale.

In addition to the non-thermal sputtering from winds, shocks from nearby supernovae might play an important role in processing dust in the filaments. While the Arches cluster might seem like a promising source of past supernova events; it is unclear whether the young cluster (2-3 Myrs; Figer et al. 2002) is old enough for the most massive stars in the cluster to have exploded as supernovae based on stellar evolution models (Meynet \& Maeder 2000; Figer 2005). Additionally, there is no evidence of a relatively recent ($\lesssim10^4$ years) supernova remnant in radio or x-ray observations of the immediate region (Yusef-Zadeh \& Morris 1988; Ponti et al. 2015). 

The Galactic center as a whole is a rich environment with many massive stars. The expected supernova rate at the Galactic center is largely dependent on clusters of stars in the local environment (e.g., Zubovas et al. 2013). It is exceedingly difficult to trace stellar populations back over very long timescales since even the largest clusters in the Galactic center are expected to evaporate over very short timescales ($\sim10$ Myr; Kim et al. 1999; Kim et al. 2000). Because of this, it is difficult to tie grain processing to past supernova event(s) in the Galactic center, but this remains an intriguing possibility.

Alternatively, we also consider the possibility that the processing of the ISM in the region is not due to stellar sources, but rather that the grain growth process in this environment may produce grains that are smaller than the dust produced in other parts of the Galaxy. There is evidence that the dust in the nearby Sickle HII region is also composed of 0.01 $\mu$m grains (Lau et al 2016), which might imply a more global effect. However, the circumnuclear disk (CND) around Sgr A* appears to have a characteristic grain size of 0.1 $\mu$m (Lau et al. 2013). Both of these results were determined using dust temperature analysis similar to that done in this work.

It is well established that the ISM in the Galactic center is warmer, denser, and more turbulent than other parts of the galaxy (Morris \& Serabyn 1996). The dense environment in the Galactic center would favor efficient grain growth (Hirashita 2000); however, the turbulence in this environment could potentially destroy grains via grain-grain collisions if the velocity differences of the grains are large enough to shatter them ($\sim2.7$ km/s for silicates and $\sim1.2$ km/s for graphite; Jones et al. 1996). Detailed models of the evolution of dust grain sizes (Hirashita \& Yan 2009) have shown that grain shattering is more efficient at destroying large grains. This can be easily understood if we consider the collisional timescale for grains, $\tau_{col}\sim(\pi a_{A}^2)^{-1} n_{B}^{-1} v_{A}^{-1}$. Where $a_{A}$ is the radius of dust grain A, $n_{B}$ is the number density of dust grains B, and $v_{A}$ is the velocity of grain A. Since $\tau_{col}\propto a_{A}^{-2}$, the timescale for collisions declines rapidly with increasing grain size. The measured velocity dispersion of the filaments' clouds is $\sim30$ km/s (LGM2001), which appears to be more than sufficient to accelerate grains above the shattering velocity. This hypothesis warrants detailed modeling of both the destruction and growth of dust grains to determine a maximum grain size based on the characteristics of the Galactic center environment.

\section{Conclusions}

In this paper, the thermal emission from dust in the Arched Filaments was studied to better understand the heating of the region and to gather evidence for how the Arches cluster is influencing the filaments. Color-temperature maps of the region show a relatively uniform temperature over the length of the filaments, which is interesting considering that the filaments extend over large spatial scales ($\sim$25 pc). The observed temperature uniformity is inconsistent with the Arches cluster heating the region when assuming the equilibrium heating of 0.1 $\mu$m silicate grains. However, this inconsistency can be solved if the characteristic grain size in this region is smaller (0.01 $\mu$m). Properties of the filaments were modeled to determine the total infrared luminosity ($L_{IR}=8.2 \pm 2.5 \times 10^6 L_{\odot}$) and dust mass ($M_d= 6.8 \pm 2.4~\mathrm{M{_{\odot}}}$). The dust-covering fraction implied by the infrared luminosity of the filaments suggests that the Arches cluster is the primary source of heating for the filaments. DustEM models of select regions in the filaments were used to show that the filaments exhibit variations in the abundances of PAHs and SamC grains. Both the variations in grain abundances and depletion of larger dust grains suggest that grain processing mechanisms have been operating in the region; however, our analysis indicates that the Arches cluster plays little role in these effects. The observed grain sizes in the Arched Filaments are similar to those in the Sickle region, potentially implying a more global effect. Supernova events may play a role in dust processing but direct linkage to past events in the Galactic center is presently infeasible. Alternatively, the smaller dust grain size may be linked to high velocity grain collisions that are capable of shattering grains. This hypothesis deserves more investigation and could significantly impact our understanding of the dust population in the Galactic center.

\emph{Acknowledgments} We would like to thank the rest of the FORCAST team, George Gull, Justin Schoenwald, Chuck Henderson, Joe Adams, the USRA Science and Mission Ops teams, and the entire SOFIA staff. We would also like to thank the anonymous referee for the useful comments and suggestions on this paper. This work is based on observations made with the NASA/DLR Stratospheric Observatory for Infrared Astronomy (SOFIA). SOFIA science mission operations are conducted jointly by the Universities Space Research Association, Inc. (USRA), under NASA contract NAS2-97001, and the Deutsches SOFIA Institut (DSI) under DLR contract 50 OK 0901. Financial support for FORCAST was provided by NASA through award 8500-98-014 issued by USRA. This research has made use of the NASA/ IPAC Infrared Science Archive, which is operated by the Jet Propulsion Laboratory, California Institute of Technology, under contract with the National Aeronautics and Space Administration. This material is based upon work supported by the National Science Foundation Graduate Research Fellowship under Grant No. DGE-1144153.

\clearpage

\vfill

\section*{Figures}

\begin{sidewaysfigure}
\centering
\includegraphics[width=105mm,scale=1.0]{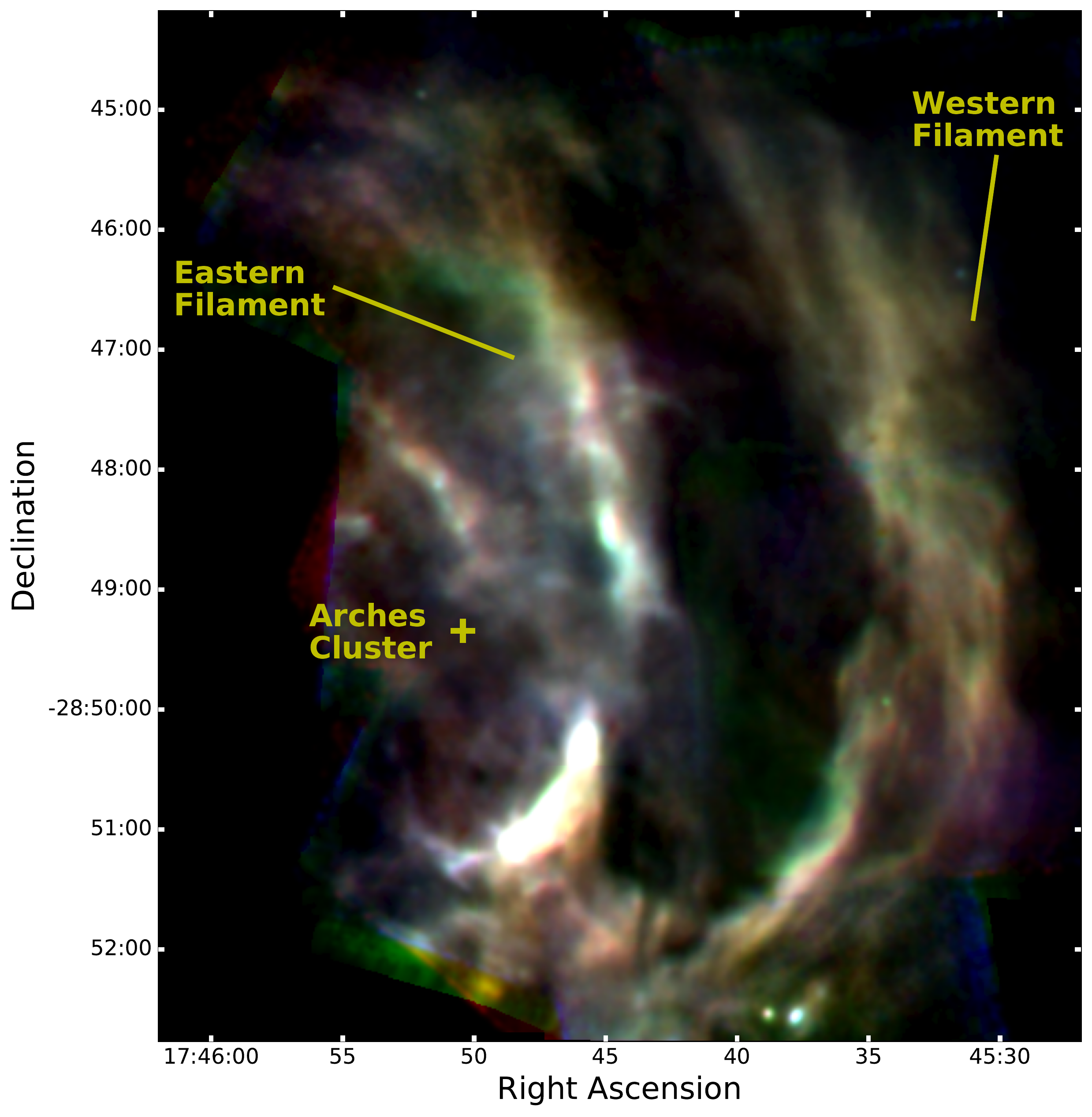}
\includegraphics[width=105mm,scale=1.0]{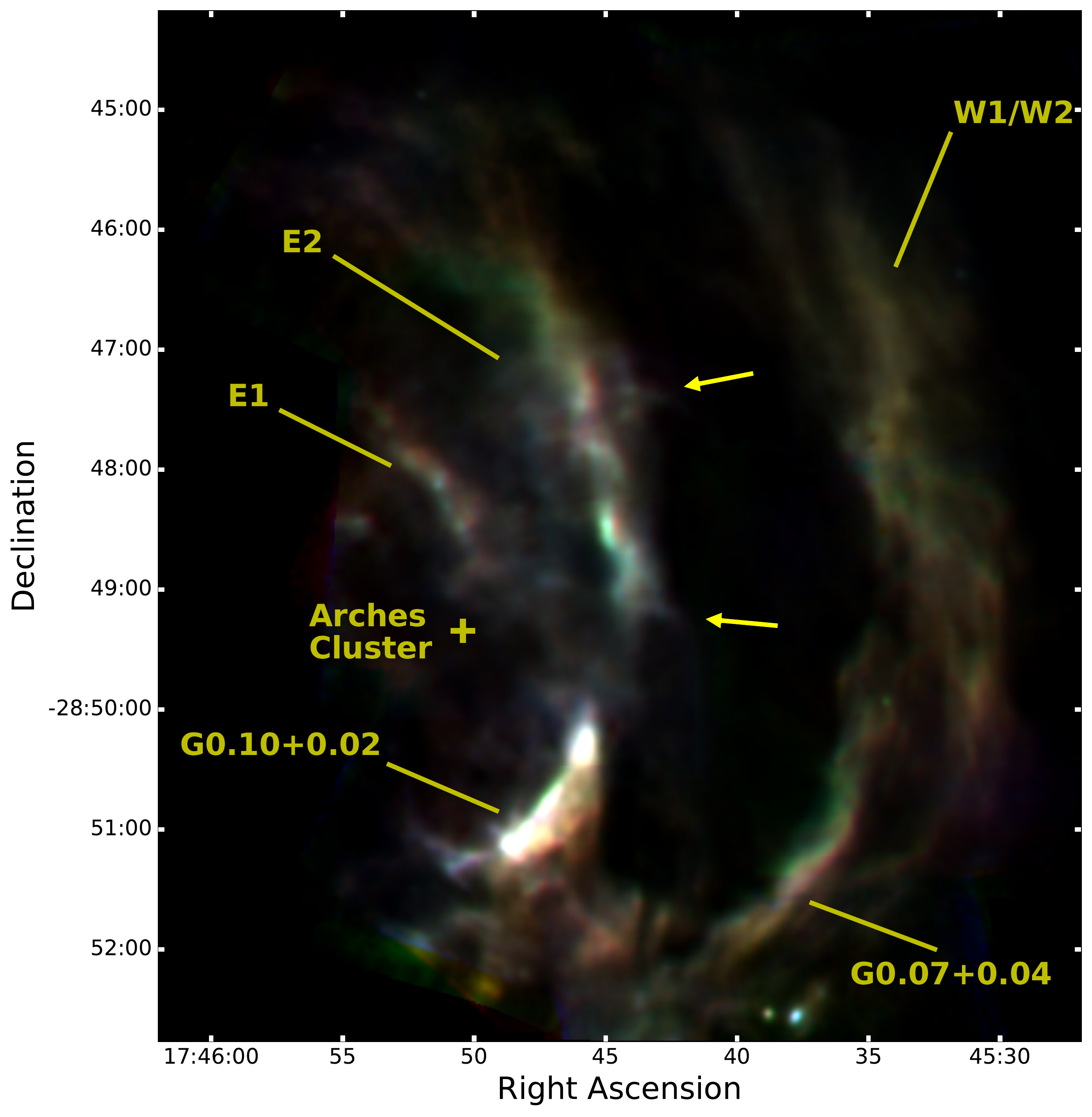}

\caption{{\footnotesize FORCAST 25.2 (blue), 31.5 (green), and 37.1 (red) $\mu$m false color images of the Arched Filaments. The images at each wavelength are a mosaic of six individual FORCAST fields. The images have been deconvolved and convolved back to a uniform beam size of 3.4". The left and right panels depict the same data, but the images have a different stretch and background cutoff to better show the bright (right) and diffuse (left) features. The maps have been labeled with the naming conventions for regions consistent with the literature. The yellow arrows mark the locations of a few wispy features that are discussed in the text. For reference, 1' at the distance of the Galactic center corresponds to $\sim$2.3 pc.}}
\label{fig:fig1}
\end{sidewaysfigure}

\begin{figure}[ht]
\centering
\includegraphics[width=120mm,scale=1.0]{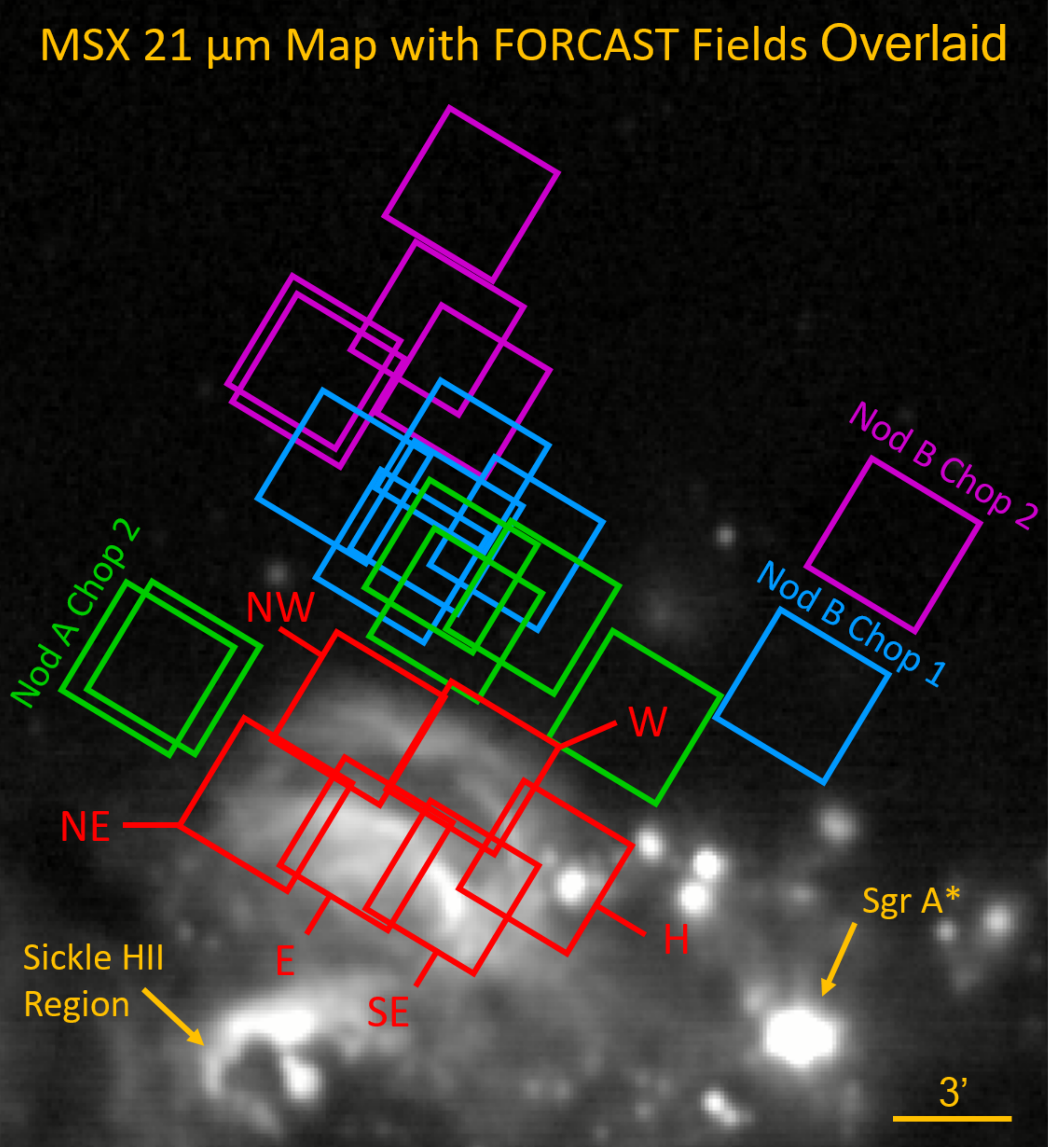}
\caption{{\footnotesize Locations of the observed FORCAST fields. The background grayscale image is the MSX 21 $\mu$m map of the region. The observed fields of the Arched Filaments are labeled and denoted in red. Because of the asymmetric chop used for the observations, there are three ``off-source'' positions for each ``on-source position.'' For each observation, nod A has one position on the source (red) and one off (green). The matching nod B has the same chop throw but both positions are located off the source (purple and blue). The locations of Sgr A* and the Sickle HII region are provided for reference.}}
\label{fig:fig2}
\end{figure}

\begin{figure}[ht]
\centering
\includegraphics[width=130mm,scale=1.0]{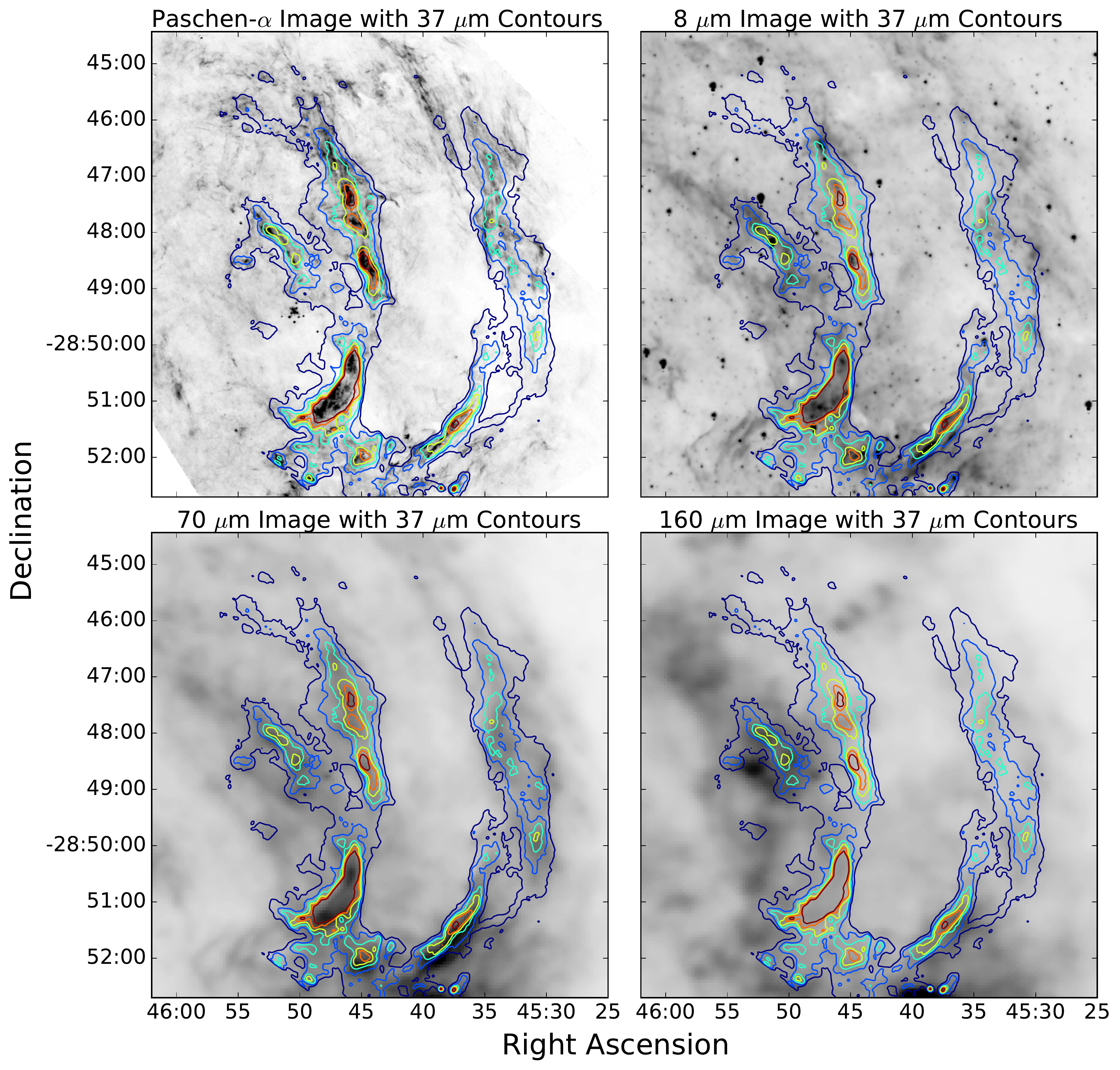}
\caption{{\footnotesize Comparison of the morphology of the FORCAST 37.1 $\mu$m emission (contours: 10, 15, 20, 25, 30, and 35-$\sigma$) with emission from Paschen-$\alpha$ (top left),  8 $\mu$m (top right), 70 $\mu$m (bottom left), and 160 $\mu$m (bottom right) in the region. The Right Ascension axis of the displayed maps shows minutes and seconds of time to be added to 17$^h$. There is relatively good spatial correlation between the warm dust traced by FORCAST and the Paschen-$\alpha$ emission that traces the HII region. Note that, the 37.1 $\mu$m emission correlates well with the 8 $\mu$m emission in some regions but follows the Paschen-$\alpha$ more closely in others. The 70 $\mu$m emission also follows the warm dust component well; however, the 160 $\mu$m emission distribution is markedly different from that at shorter wavelengths.}}
\label{fig:fig8}
\end{figure}

\newpage

\begin{sidewaysfigure}
\centering
\includegraphics[width=80mm,scale=1.0]{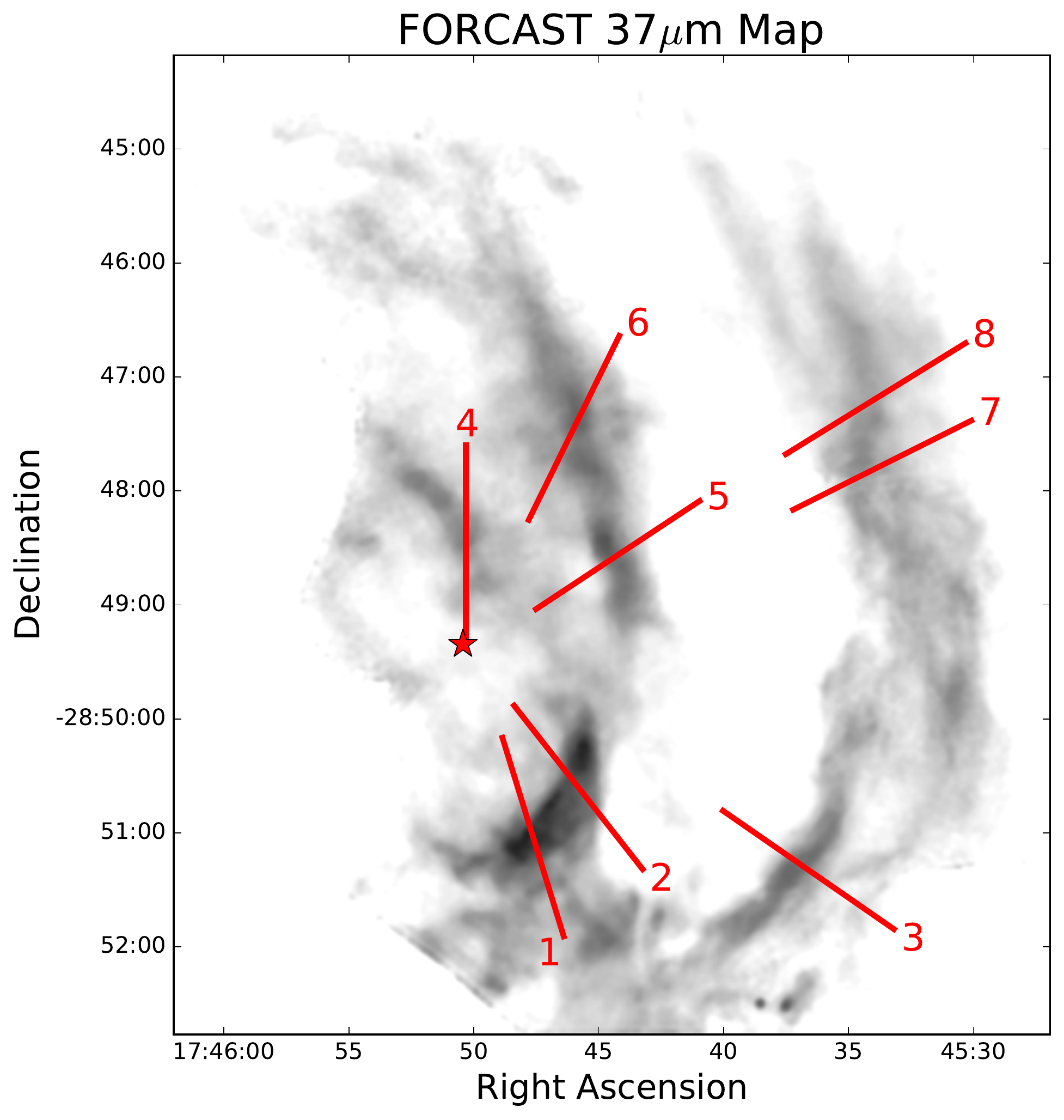}
\includegraphics[width=120mm,scale=1.0]{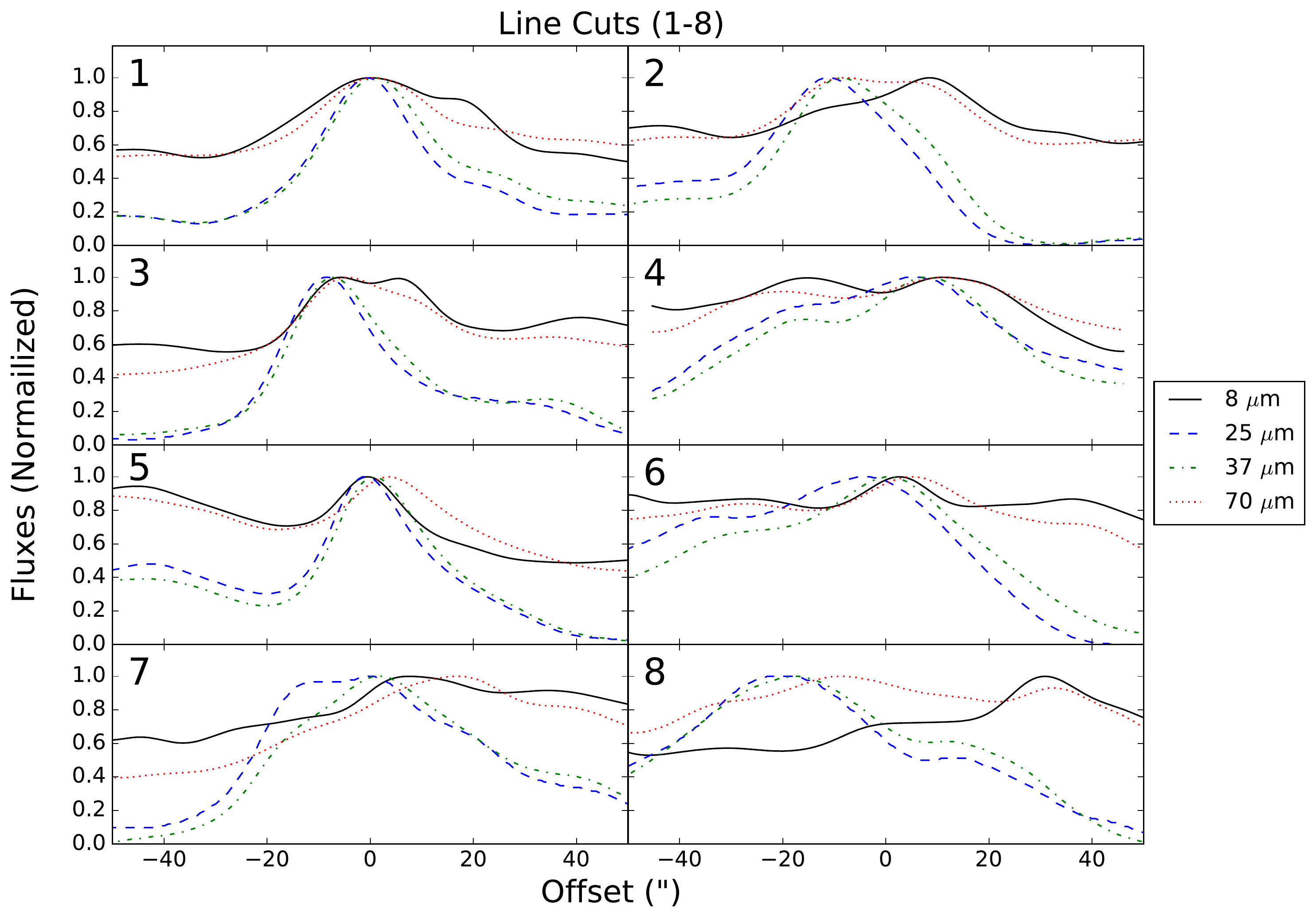}
\caption{{\footnotesize Left: FORCAST 25$\mu$m map of the Arched Filaments with positions of the linecuts indicated. Right: Plots of linecuts through the filaments at various positions. The 8, 25.2, 37.1 and 70 $\mu$m data are plotted in black, blue, green, and red, respectively. The directions of the lines run from negative to positive values with increasing distance from the  position of the Arches cluster (marked by the star). Lines 2 and 8 show a significant emission peak at 8 $\mu$m that is spatially offset from the main emission peaks at FORCAST wavelengths. The additional 8 $\mu$m peaks are also traced by the 70 $\mu$m, indicating that the emission likely originates relatively cool dust in the clouds.}}
\label{fig:fig9}
\end{sidewaysfigure}

\begin{figure}[ht]
\centering
\includegraphics[width=150mm,scale=1.0]{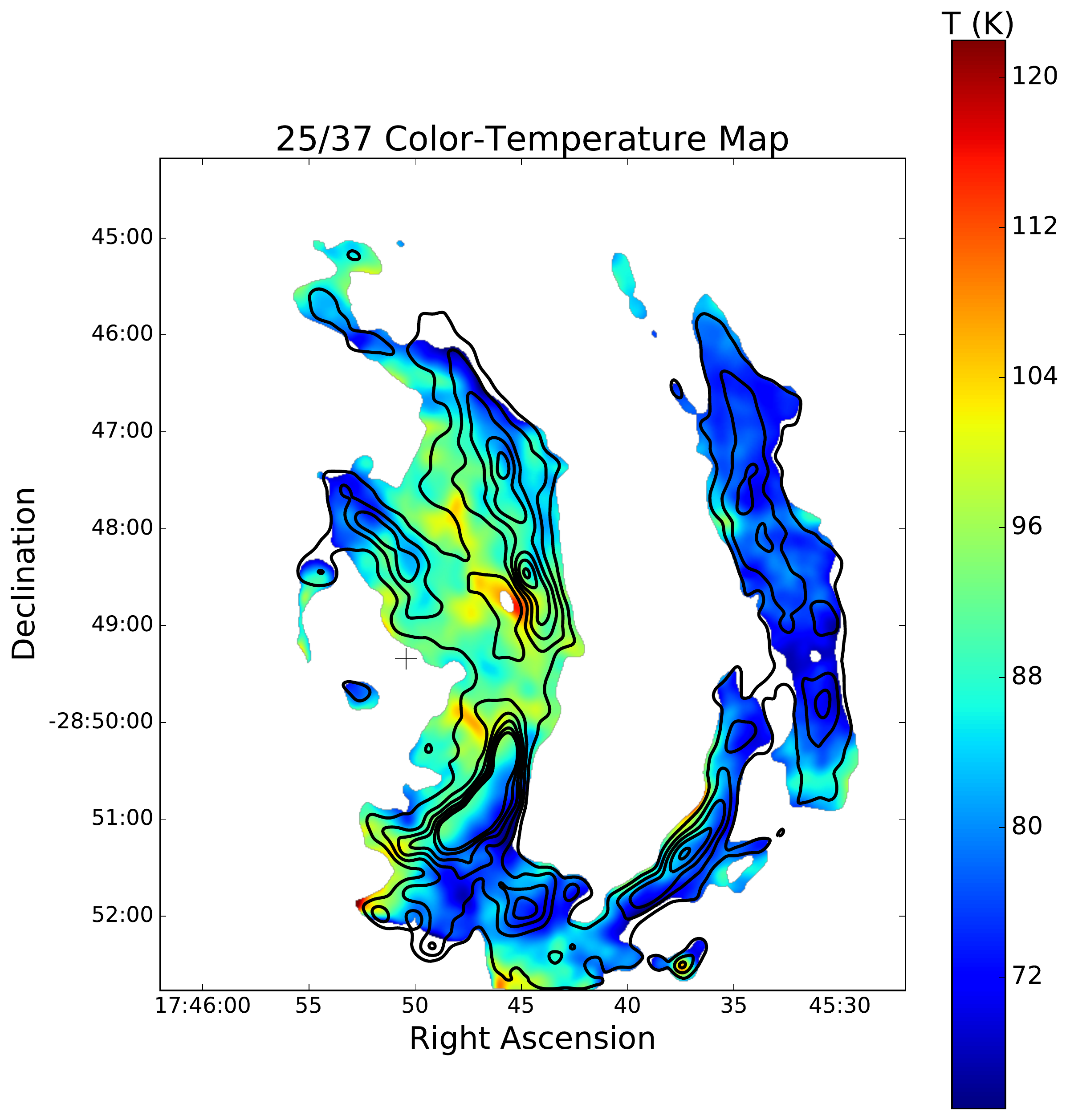}
\caption{{\footnotesize Color-Temperature map of the Arches using the 25.2 and 37.1 $\mu$m FORCAST maps of the Arched Filaments. The map has been smoothed using a Gaussian kernel with a FWHM of 7". The 37.1 $\mu$m contours from 0.2 to 0.7 Jy in increments of 0.1 Jy have been overlaid on the figure for reference. The position of the Arches cluster has been marked with a cross.}}
\label{fig:fig3}
\end{figure}

\begin{sidewaysfigure}
\centering
\includegraphics[width=100mm,scale=1.0]{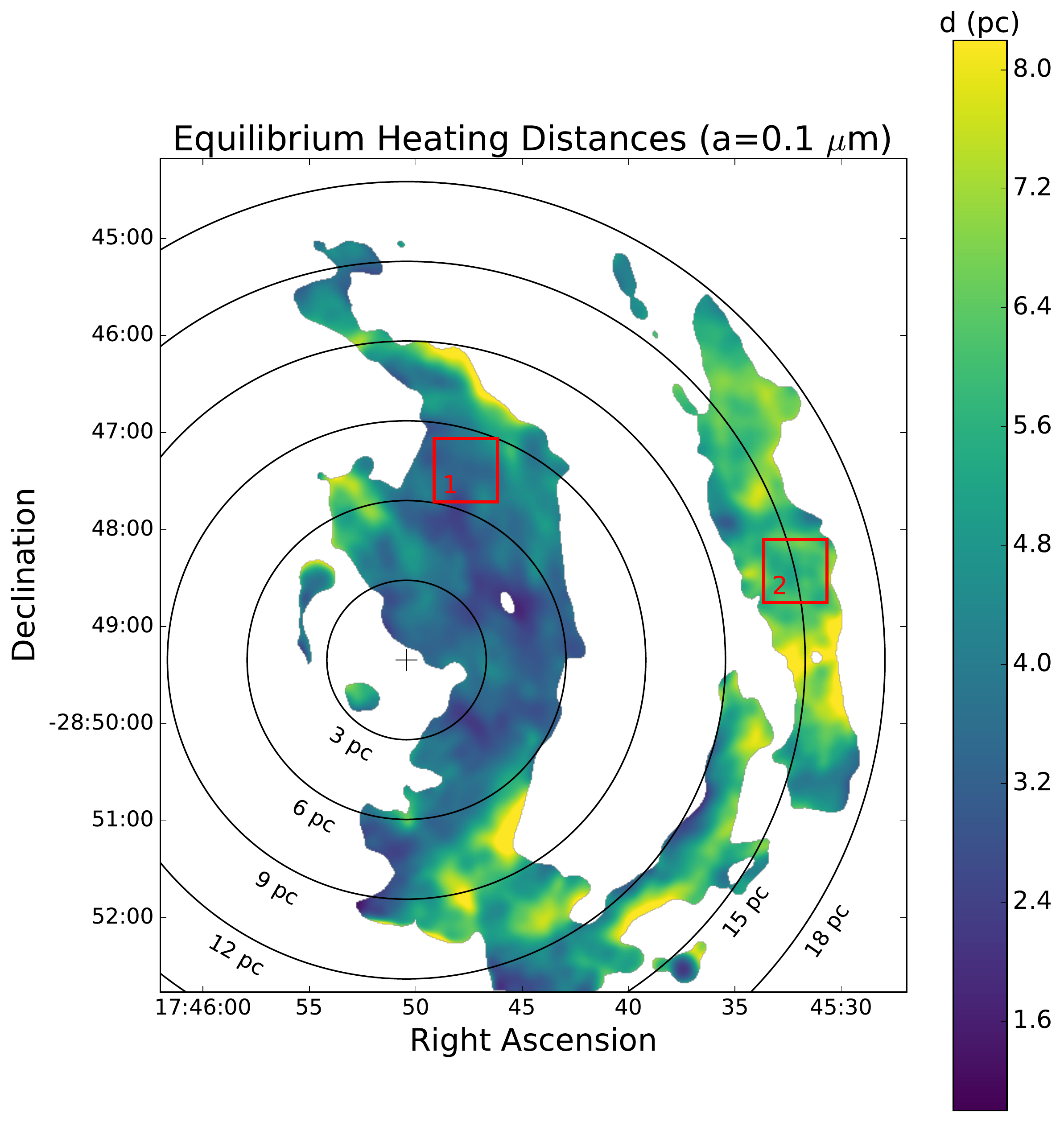}
\includegraphics[width=100mm,scale=1.0]{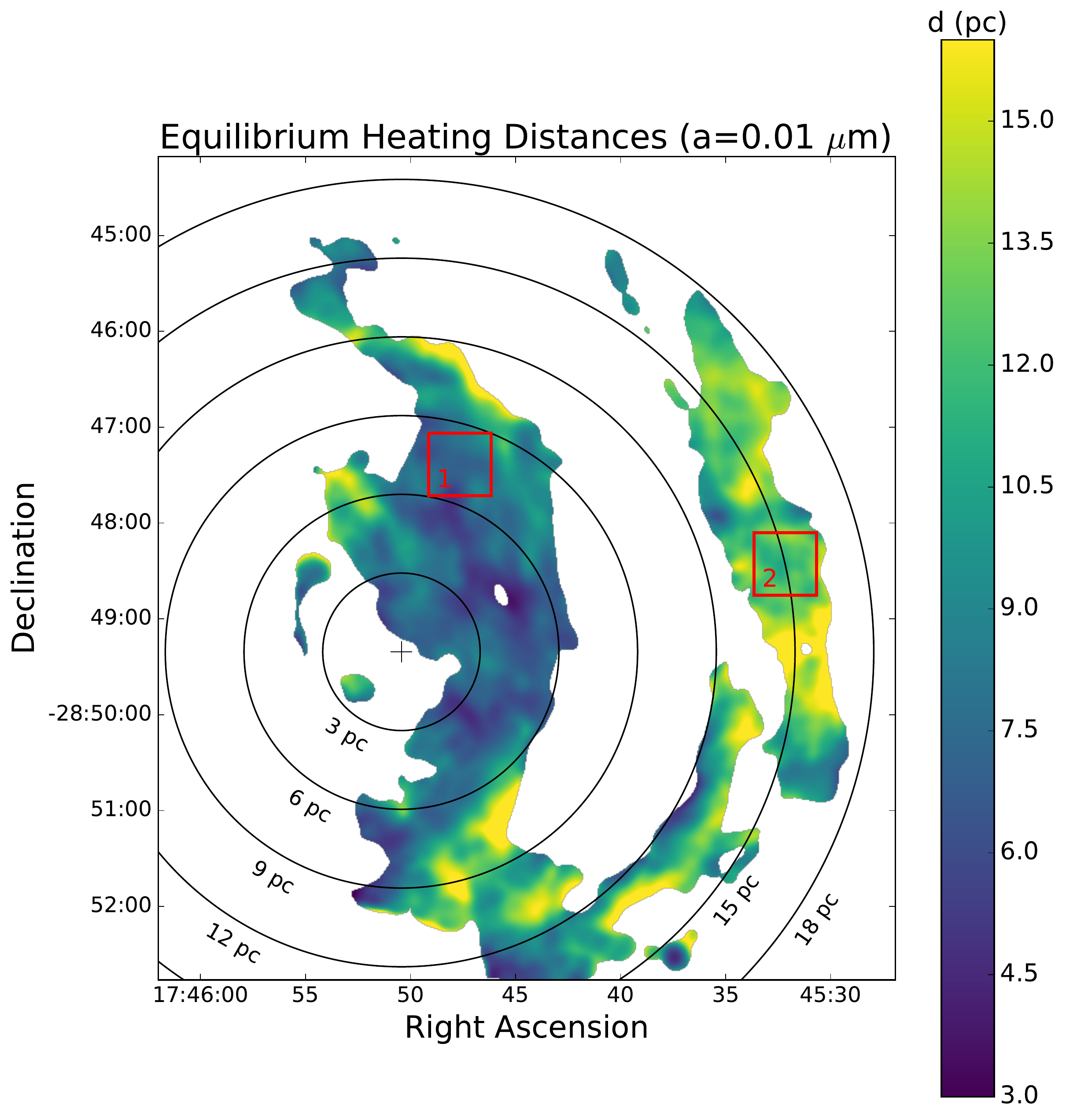}
\caption{{\footnotesize Derived equilibrium distances between the Arched Filaments and the Arches cluster assuming the cluster is the heating source for the filaments. The color maps indicate the calculated equilibrium heating distance from the Arches cluster, indicated on the associated color bar for each figure. The maps in the left and right panels indicate the distances implied by the heating of 0.1 $\mu$m and 0.01 $\mu$m dust grains, respectively. The location of the Arches cluster is marked with a cross. Circles centered on the cluster indicate projected distances from the cluster in increments of 3 pc from 3 to 18 pc. The red boxes mark two locations described in the text.}}
\label{fig:fig4}
\end{sidewaysfigure}

\begin{sidewaysfigure}
\centering
\includegraphics[width=100mm,scale=1.0]{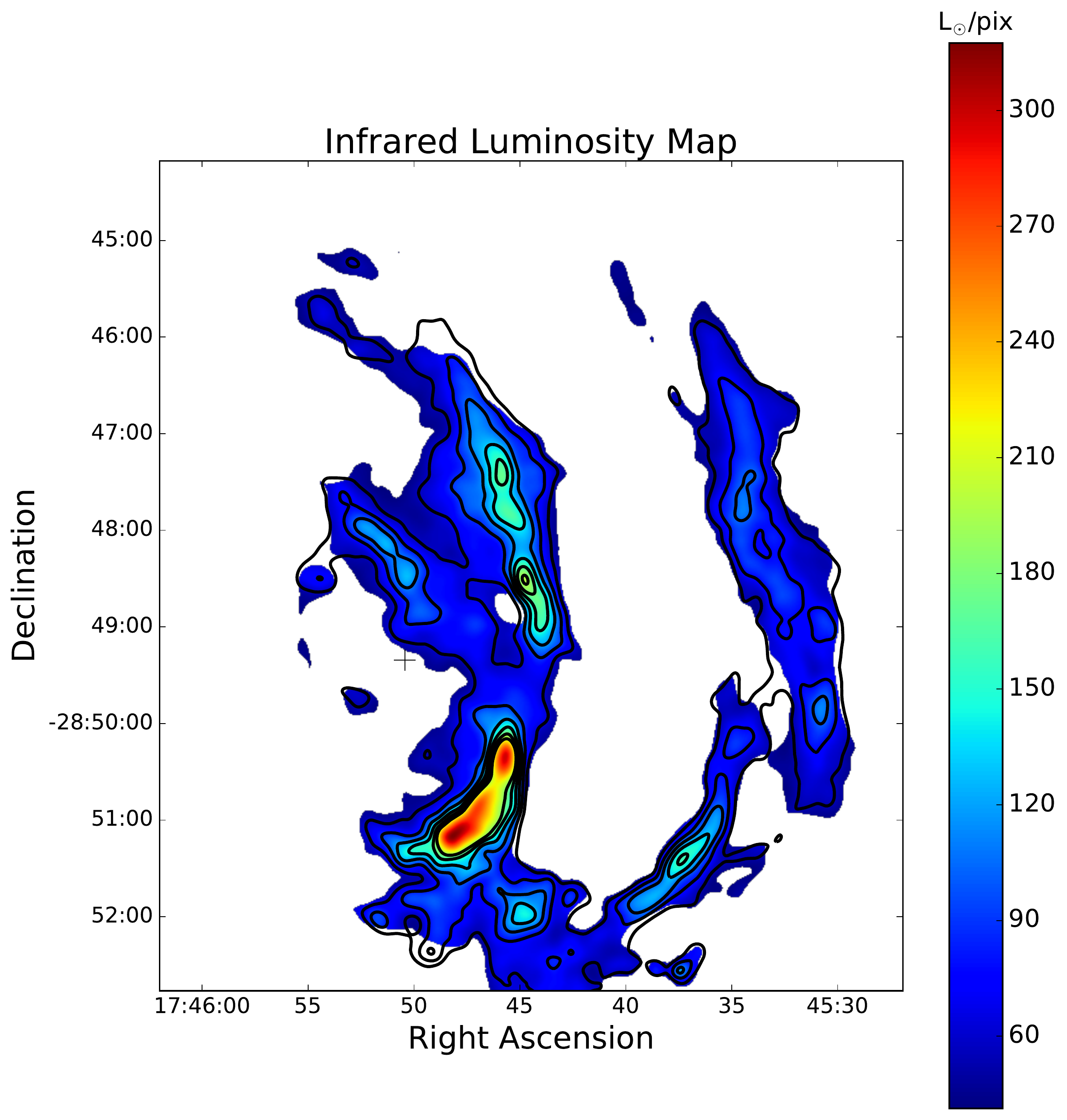}
\includegraphics[width=100mm,scale=1.0]{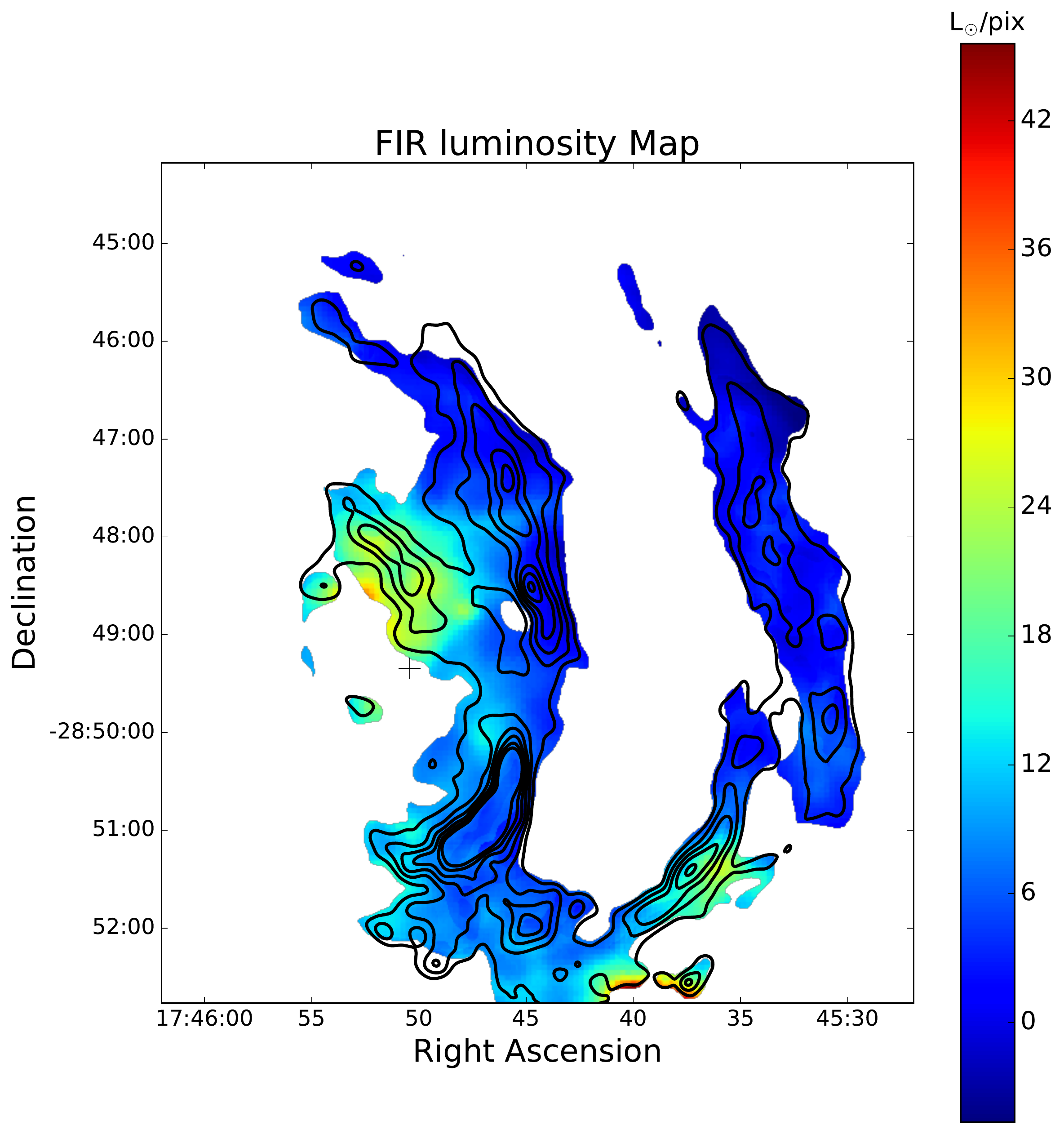}
\caption{{\footnotesize Luminosity maps of the filaments in L$_{\odot}$/pixel with a pixel size of 0.768"$\times$0.768". The maps have been separated into two components. The warm dust traced by FORCAST is presented on the left. The cold dust component traced by the far-infrared emission is presented on the right. The 37.1 $\mu$m contours from 0.2 to 0.7 Jy in increments of 0.1 Jy have been overlaid on both figures for reference. The location of the Arches cluster is marked by a cross. The most luminous region in the filaments is G0.10+0.02 with L$_{IR}=8^{+2.0}_{-1.5}\times10^5 L_{\odot}$ contained within the top three contours.}}

\label{fig:fig5}
\end{sidewaysfigure}

\begin{figure}[ht]
\centering
\includegraphics[width=150mm,scale=1.0]{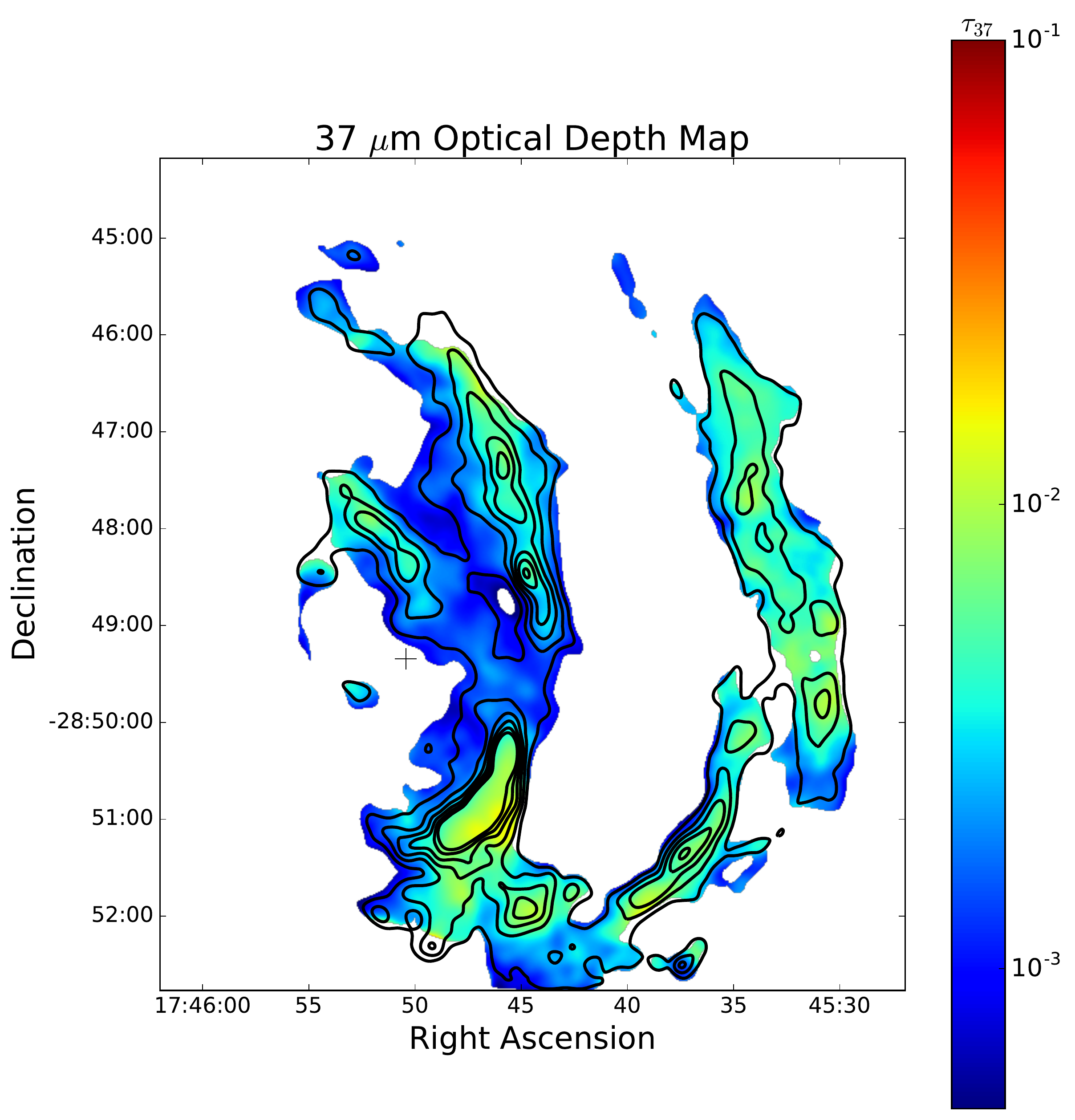}
\caption{{\footnotesize 37.1 $\mu$m optical depth map of the Arched Filaments region. The 37.1 $\mu$m contours from 0.2 to 0.7 Jy in increments of 0.1 Jy have been overlaid on the figure for reference. The location of the Arches cluster is marked by a cross.}}
\label{fig:fig6}
\end{figure}

\begin{sidewaysfigure}
\centering
\includegraphics[width=80mm,scale=1.0]{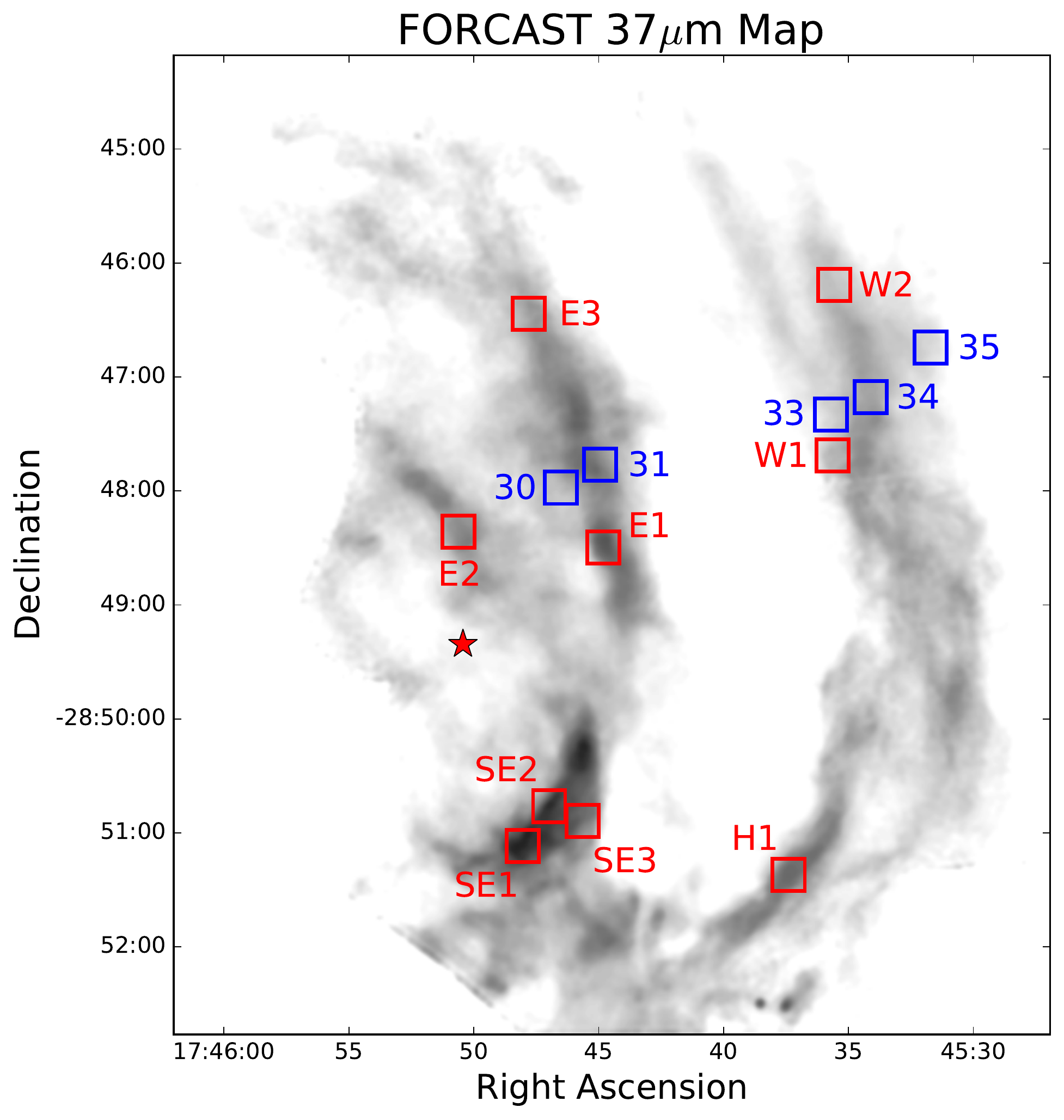}
\includegraphics[width=120mm,scale=1.0]{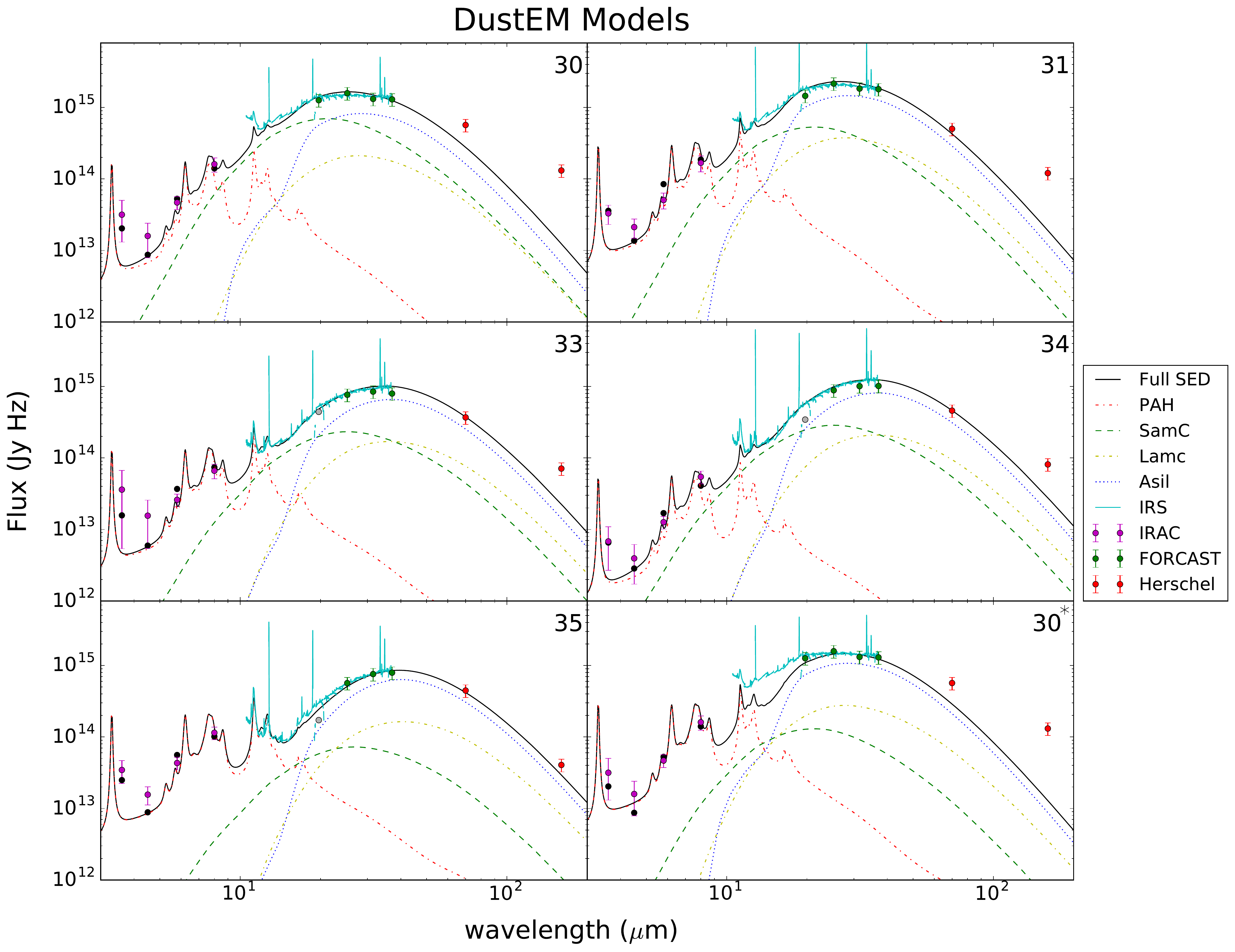}
\caption{{\footnotesize Left: Regions of the Arched Filaments modeled with DustEM. Regions with Spitzer/IRS spectra are shown in blue, while other regions that were modeled are marked in red. Right: Best-fit models for the regions with Spitzer/IRS spectra. Models for the red regions are shown in Figure 10. Detailed parameters for models can be found in Table 3. The grey points in the plots indicate data that were not used in the fitting process because of possible issues with the flux calibration. We also plot the 160 $\mu$m data to show any far-infrared excess though it is not used in the fitting process. The second model of Position 30 located at bottom right uses the standard SamC abundance to show the discrepancy of the model and the observed continuum from 10-20$\mu$m.}}
\label{fig:fig6}
\end{sidewaysfigure}

\newpage

\begin{sidewaysfigure}
\includegraphics[width=210mm,scale=1.0]{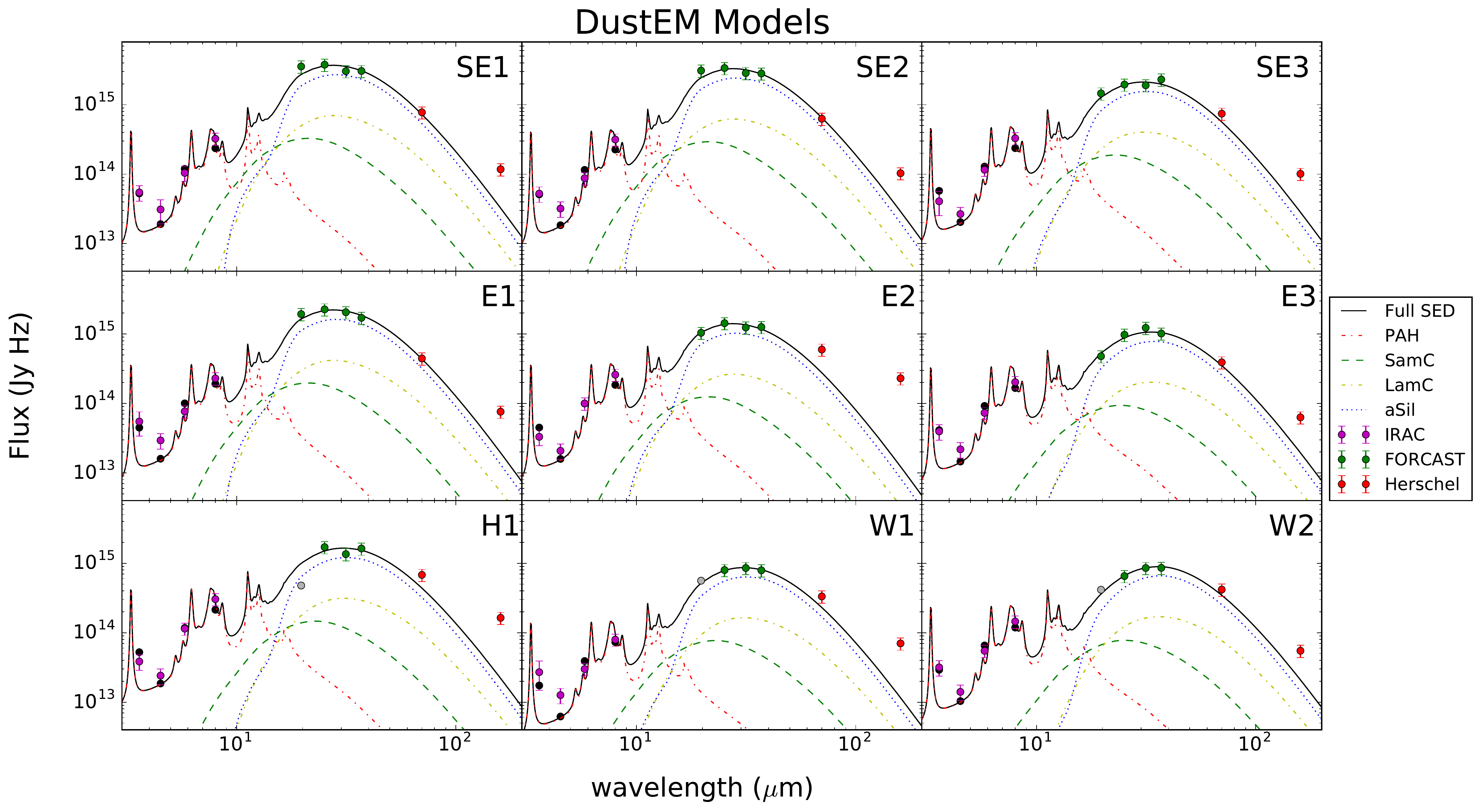}
\caption{{\footnotesize Additional regions of the Arched Filaments modeled with DustEM. Locations of regions are shown in Figure 9 as red squares. Best-fit model parameters can be found in Table 3. The grey points in the plots indicate data that were not used in the fitting process because of possible issues with the flux calibration. We also plot the 160 $\mu$m data to show any far-infrared excess though it is not used in the fitting process.}}
\end{sidewaysfigure}

\section*{Tables}

\begin{table}[htp]
\footnotesize
\centering
\caption{\ Observation Details}
\label{tab:flightDetails}
\begin{tabular}{cccccccc}
\hline
 Cycle  & Flight & Date & Altitude (ft.) & Field & Filter 1 ($\mu$m) & Filter 2 ($\mu$m) & Integration (s)\\
\hline
\hline
3 & 217 & 6/13/15 & $\sim$36,000 & H & 19.7 & 31.5 & 337\\
3 & 217 & 6/13/15 & $\sim$36,000 & H & 25.2 & 31.5 & 301\\
3 & 217 & 6/13/15 & $\sim$36,000 & H & N/A & 37.1 & 345\\

3 & 224 & 7/3/15 & $\sim$38,000 & NW & 19.7 & 37.1 & 263\\
3 & 224 & 7/3/15 & $\sim$38,000 & NW & 25.2 & 31.5 & 285\\
3 & 224 & 7/3/15 & $\sim$38,000 & NW & N/A & 37.1 & 141\\

3 & 225 & 7/4/15 & $\sim$38,000 & NE & 19.7 & 37.1 & 52\\
3 & 225 & 7/4/15 & $\sim$38,000 & NE & 25.2 & 31.5 & 166\\
3 & 225 & 7/4/15 & $\sim$38,000 & NE & N/A & 37.1 & 54\\

3 & 225 & 7/4/15 & $\sim$38,000 & E & 19.7 & 37.1 & 168\\
3 & 225 & 7/4/15 & $\sim$38,000 & E & 25.2 & 31.5 & 213\\
3 & 225 & 7/4/15 & $\sim$38,000 & E & N/A & 37.1 & 173\\

3 & 225 & 7/4/15 & $\sim$38,000 & W & 25.2 & 37.1 & 248\\

3 & 227 & 7/7/15 & $\sim$39,000 & SE & 19.7 & 37.1 & 210\\
3 & 227 & 7/7/15 & $\sim$39,000 & SE & 25.2 & 31.5 & 222\\
3 & 227 & 7/7/15 & $\sim$39,000 & SE & N/A & 37.1 & 212\\

4 & 321 & 7/14/16 & $\sim$43,000 & W & 19.7 & 31.5 & 773\\

4 & 324 & 7/19/16 & $\sim$38,000 & NE & 19.7 & 37.1 & 792\\
\hline
\end{tabular}
\end{table}

\begin{table}[htp]
\footnotesize
\centering
\caption{\ Starburst99 Model Parameters}
\label{tab:S99models}
\begin{tabular}{cc}

\hline
Parameter & Value \\ 
\hline
\hline

Starbust Type & Fixed Mass \\  
Stellar Mass & $7\times10^4 M_{\odot}$ \\
IMF Law & Kroupa \\
IMF Exponents  & 1.3, 2.3 \\
Mass Ranges & 0.1-0.5, 0.5-120 $M_{\odot}$ \\
Age & 2.5 Gyr \\
Metalicity & 0.040 \\
Evolution Tracks & Geneva v00 \\
Stellar Atmospheres & Pauldrach/Hillier \\
Wind Model & Evolution \\

\hline
\hline
\end{tabular}
\\Model Parameters used to simulate the radiation field from the Arches cluster. 
\centering
\end{table}

\begin{table}[htp]
\footnotesize
\centering
\caption{\ DustEM Model Parameters}
\label{tab:DustEMmodels}
\begin{tabular}{ccccccc}

\hline
Location & aSil & Column Density (g/cm$^2$) & $X_{PAH}$ & $X_{SamC}$ &  $d$ (pc) & $\chi^2$ \\ 
\hline
\hline

30 & a=0.01 $\mu$m & $2.5 ^{+6.1}_{-2.2} \times10^{-5}$ & $0.29^{+0.11}_{-0.10}$ & 200.0 & $6.0^{+3.7}_{-3.5}$ & 1.6 \\  
31 & a=0.01 $\mu$m & $3.2 ^{+2.3}_{-1.5} \times10^{-5}$ & $0.23^{+0.09}_{-0.09}$ & 80.0 & $5.8 ^{+1.5}_{-1.5}$ & 1.2 \\ 

33 & a=0.01 $\mu$m & $3.2 ^{+3.6}_{-2.0} \times10^{-5}$ & $0.29^{+0.08}_{-0.08}$ & 20.0 & $8.8 ^{+2.7}_{-2.9}$ & 0.8 \\
34 & a=0.01 $\mu$m & $4.9 ^{+4.0}_{-2.7} \times10^{-5}$ & $0.11^{+0.05}_{-0.06}$ & 20.0 & $9.8 ^{+2.4}_{-2.6}$ & 1.6 \\  
35 & a=0.01 $\mu$m & $6.5 ^{+4.9}_{-3.6} \times10^{-5}$ & $0.51^{+0.14}_{-0.10}$ & 10.0 & $12.7 ^{+2.6}_{-3.2}$ & 0.9 \\
35$^{\dagger}$ & a=0.01 $\mu$m & $9.1 ^{+4.2}_{-3.4} \times10^{-5}$ & $0.38^{+0.11}_{-0.11}$ & 10.0 & $16.5 ^{+2.1}_{-2.3}$ & 0.9 \\

SE1 &  a=0.01 $\mu$m & $4.1^{+4.6}_{-2.8}\times10^{-5}$ & $0.24^{+0.10}_{-0.10}$ & x & $5.5^{+1.9}_{-2.1}$ & 1.7 \\  
SE2 & a=0.01 $\mu$m & $3.5^{+2.8}_{-1.9}\times10^{-5}$ & $0.26^{+0.10}_{-0.11}$  & x & $5.0^{+1.4}_{-1.5}$ & 1.7 \\ 
SE3 & a=0.01 $\mu$m & $6.5^{+7.6}_{-4.6}\times10^{-5}$ & $0.40^{+0.19}_{-0.19}$  & x & $8.0^{+2.7}_{-3.2}$ & 2.1 \\

E1 & a=0.01 $\mu$m & $2.5^{+1.6}_{-1.2}\times10^{-5}$ & $0.38^{+0.13}_{-0.13}$ & x & $5.1^{+1.2}_{-1.3}$ & 1.1 \\  
E2 & a=0.01 $\mu$m & $3.0^{+4.4}_{-2.1}\times10^{-5}$ & $0.41^{+0.22}_{-0.22}$  & x & $6.7^{+2.8}_{-2.8}$ & 2.9 \\  
E3 & a=0.01 $\mu$m & $4.2^{+1.4}_{-1.1}\times10^{-6}$ & $0.55^{+0.19}_{-0.20}$  & x & $8.8^{+1.0}_{-1.0}$ & 1.3 \\  

W1 & a=0.01 $\mu$m & $2.9^{+1.8}_{-1.3}\times10^{-5}$ & $0.35^{+0.11}_{-0.11}$  & x & $8.3^{+1.6}_{-1.8}$ & 1.1 \\
W2 & a=0.01 $\mu$m & $5.7^{+2.3}_{-1.9}\times10^{-5}$ & $0.64^{+0.19}_{-0.19}$  & x & $11.5^{+1.5}_{-1.6}$ & 1.5 \\ 
H1 & a=0.01 $\mu$m & $3.9^{+5.8}_{-3.3}\times10^{-5}$ & $0.39^{+0.21}_{-0.21}$  & x & $7.0^{+2.9}_{-3.3}$ & 3.1 \\ 

\hline

30 & $dn/da\propto a^{-3.4}$ & $3.3^{+5.4}_{-2.5}\times10^{-5}$ & $0.08^{+0.01}_{-0.03}$ & 15.0 & $3.7^{+1.8}_{-1.8}$  & 1.5 \\ 
31 & $dn/da\propto a^{-3.4}$ & $3.6^{+2.0}_{-1.5}\times10^{-5}$ & $0.07^{+0.02}_{-0.03}$ & 8.0 & $3.3^{+0.7}_{-0.7}$ & 1.3 \\ 

33 & $dn/da\propto a^{-3.4}$ & $4.0^{+2.4}_{-1.7} \times10^{-5}$ & $0.08^{+0.02}_{-0.02}$ & 4.0 & $5.4^{+1.1}_{-1.1}$ & 0.8 \\
34 &  $dn/da\propto a^{-3.4}$ &$5.7^{+2.5}_{-2.0} \times10^{-5}$ & $0.03^{+0.01}_{-0.02}$ & 4.0 & $5.9^{+0.9}_{-0.9}$ & 1.7 \\
35 &  $dn/da\propto a^{-3.4}$ &$8.3^{+2.9}_{-2.7} \times10^{-5}$ & $0.14^{+0.04}_{-0.04}$ & 1.0 & $7.9^{+0.9}_{-1.0}$ & 0.9 \\
35$^{\dagger}$ & $dn/da\propto a^{-3.4}$ & $1.3^{+0.3}_{-0.3} \times10^{-4}$ & $0.11^{+0.03}_{-0.03}$ & 1.0 & $10.9^{+0.8}_{-0.9}$ & 1.0 \\

SE1 &  $dn/da\propto a^{-3.4}$ & $4.2^{+3.2}_{-2.0}\times10^{-5}$ & $0.07^{+0.03}_{-0.03}$ & x & $2.8^{+0.8}_{-0.8}$ & 1.7 \\  
SE2 & $dn/da\propto a^{-3.4}$ & $3.8^{+1.2}_{-0.9}\times10^{-5}$ & $0.07^{+0.03}_{-0.03}$  & x & $2.8^{+0.4}_{-0.5}$ & 1.9 \\  
SE3 & $dn/da\propto a^{-3.4}$ & $7.2^{+5.2}_{-3.4}\times10^{-5}$ & $0.11^{+0.06}_{-0.06}$  & x & $4.7^{+1.1}_{-1.1}$ & 2.2 \\

E1 & $dn/da\propto a^{-3.4}$ & $2.7^{+0.7}_{-0.7}\times10^{-5}$ & $0.11^{+0.03}_{-0.04}$ & x & $2.9^{+0.3}_{-0.4}$ & 1.2 \\  
E2 & $dn/da\propto a^{-3.4}$ & $3.9^{+6.4}_{-2.9}\times10^{-5}$ & $0.11^{+0.06}_{-0.06}$  & x & $4.2^{+2.0}_{-1.9}$ & 3.3 \\  
E3 & $dn/da\propto a^{-3.4}$ & $4.9^{+3.1}_{-2.1}\times10^{-5}$ & $0.16^{+0.05}_{-0.06}$  & x & $5.3^{+1.1}_{-1.1}$ & 1.7 \\ 

W1 & $dn/da\propto a^{-3.4}$ & $3.3^{+1.8}_{-1.3}\times10^{-5}$ & $0.10^{+0.03}_{-0.03}$  & x & $4.9^{+0.9}_{-1.0}$ & 1.1 \\ 
W2 & $dn/da\propto a^{-3.4}$ & $6.7^{+1.8}_{-1.5}\times10^{-5}$ & $0.16^{+0.05}_{-0.04}$  & x & $6.7^{+0.6}_{-0.6}$ & 1.0 \\ 
H1 & $dn/da\propto a^{-3.4}$ & $4.9^{+8.3}_{-4.1}\times10^{-5}$ & $0.11^{+0.06}_{-0.06}$  & x & $4.2^{+2.0}_{-2.4}$ & 3.5 \\

\hline
\hline
\end{tabular}
\\The best-fit model parameters for the regions modeled with DustEM are listed in this table. The table is split according to the type of silicate grain size distribution used in the model. The top half uses a uniform grain size with $a=0.01~\mu$m, while the bottom half uses a power law distribution of silicate grain sizes. Columns 4 \& 5 list the factors that the relative abundances of PAHs and SamC grains have been scaled to compared to the ISM abundances listed in the text. The change in the SamC abundance represents the minimum increase needed to match the observed continuum from 10-20 $\mu$m. Locations where the SamC abundance was not fit are indicated with an x. Distances to the Arches cluster implied by the models are also noted. Model 35$^{\dagger}$ is the best-fit model for location 35 using a different extinction correction for the source flux, as described in the text.

\centering

\end{table}

\end{document}